\documentclass[12pt,a4paper]{amsart}

\usepackage{amsmath,amsthm,amsfonts,amssymb,srcltx,empheq}
\usepackage{mathrsfs}
\usepackage[dvipdfmx]{graphicx}
\usepackage[all]{xy}
\usepackage{color}

\usepackage{here}
\usepackage{caption}

\newcommand{\e}{\mathrm{e}}

\def\IN{\mathbb {N}}
\def\IZ{\mathbb {Z}}

\def\IC{\mathbb {C}}

\def\br{\boldsymbol{r}}
\def\bk{\boldsymbol{k}}

\def\bs{\boldsymbol{s}}
\def\bm{\boldsymbol{m}}

\def\bel{\boldsymbol{\ell}}

\def\bsig{\boldsymbol{\sigma}}

\def\bdel{\boldsymbol{\delta}}
\def\bY{\boldsymbol{Y}}

\def\bN{\boldsymbol{N}}

\def\fc{\mathfrak{c}}
\def\fq{\mathfrak{q}}
\def\ft{\mathfrak{t}}

\def\sfs{\mathsf{s}}

\def\bft{\boldsymbol{\mathfrak{t}}}
\def\bfc{\boldsymbol{\mathfrak{c}}}

\def\ll{\left\lgroup}
\def\rr{\right\rgroup}

\DeclareMathOperator{\partit}{par}

\newcommand{\dimfin}{{M}}    
\newcommand{\dimaff}{{M}}    
\newcommand{\levelaff}{{m}}    
\newcommand{\indfin}{\overline{\mathcal I}_\dimfin}
\newcommand{\indaff}{{\mathcal I}_\dimaff}

\newcommand{\inner}[2]{\left\langle{#1},{#2}\right\rangle}

\newcommand{\slchap}[1]{\widehat{\mathfrak{sl}}(#1)}

\theoremstyle{plain}
  \newtheorem{prob}{Problem}[section]
  \newtheorem{conj}[prob]{Conjecture}

    \newtheorem{defi}[prob]{Definition}
\theoremstyle{remark}
  \newtheorem{remark}[prob]{\bf Remark}

\newtheorem{exam}[prob]{\bf Example}

\usepackage[scale=0.84]{geometry}

\makeatletter
\def\Left#1#2\Right{\begingroup%
   \def\ts@r{\nulldelimiterspace=0pt \mathsurround=0pt}%
   \let\@hat=#1%
   \def\sht@im{#2}%
   \def\@t{{\mathchoice{\def\@fen{\displaystyle}\k@fel}%
          {\def\@fen{\textstyle}\k@fel}%
          {\def\@fen{\scriptstyle}\k@fel}%
          {\def\@fen{\scriptscriptstyle}\k@fel}}}%
   \def\g@rin{\ts@r\left\@hat\vphantom{\sht@im}\right.}%
   \def\k@fel{\setbox0=\hbox{$\@fen\g@rin$}\hbox{%
      $\@fen \kern.3875\wd0 \copy0 \kern-.3875\wd0%
      \llap{\copy0}\kern.3875\wd0$}}%
      \def\pt@h{\mathopen\@t}\pt@h\sht@im%
      \Right}%
\def\Right#1{\let\@hat=#1%
   \def\st@m{\mathclose\@t}%
   \st@m\endgroup}
\makeatother

\makeatletter
 \renewcommand{\theequation}{%
       \thesection.\arabic{equation}}
 \@addtoreset{equation}{section}
\makeatother

\makeatletter
\def\eqnarray{%
 \stepcounter{equation}%
 \let\@currentlabel=\theequation
 \global\@eqnswtrue
 \global\@eqcnt\z@
 \tabskip\@centering
 \let\\=\@eqncr
 $$\halign to \displaywidth\bgroup\@eqnsel\hskip\@centering
 $\displaystyle\tabskip\z@{##}$&\global\@eqcnt\@ne
 \hfil$\displaystyle{{}##{}}$\hfil
 &\global\@eqcnt\tw@$\displaystyle\tabskip\z@{##}$\hfil
 \tabskip\@centering&\llap{##}\tabskip\z@\cr}
\makeatother

\begin{document}

\title[]{$n$-th parafermion $\mathcal{W}_N$ characters from 
\\
$U(N)$ instanton counting on ${\mathbb {C}}^2/{\mathbb {Z}}_n$
}

\author[]{Masahide Manabe} 

\address{
School of Mathematics and Statistics, 
University of Melbourne, 
Royal Parade, Parkville, Victoria 3010, Australia
}

\dedicatory{
Dedicated to the memory of Professor Omar Foda
}

\email{masahidemanabe@gmail.com}

\begin{abstract}
We propose, following the AGT correspondence, 
how the $\mathcal{W}^{\, para}_{N, n}$ 
($n$-th parafermion $\mathcal{W}_N$) minimal model characters are 
obtained from the $U(N)$ instanton counting on 
${\mathbb {C}}^2/{\mathbb {Z}}_n$ with $\Omega$-deformation 
by imposing specific conditions which remove the minimal model null states.
\end{abstract}

\maketitle

\section{Introduction}

\subsection{AGT correspondence}

The AGT correspondence \cite{Alday:2009aq} with various generalizations 
makes the connection between 
4D supersymmetric gauge theory with $\Omega$-deformation \cite{Nekrasov:2002qd} 
and 2D conformal field theory (CFT) with a generic central charge. 
In this paper 
we will focus on the correspondence between 
a 4D $\mathcal{N}=2$ $U(N)$ supersymmetric gauge theory 
on ${\IC}^2/{\IZ}_n$ and a 2D CFT with the symmetry algebra
$$
\mathcal{A}(N,n;p)=\mathcal{H} \oplus 
\widehat{\mathfrak{sl}}(n)_N \oplus 
\frac{\widehat{\mathfrak{sl}}(N)_n \oplus 
\widehat{\mathfrak{sl}}(N)_{p-N}}{\widehat{\mathfrak{sl}}(N)_{n+p-N}},
$$
which acts on the equivariant cohomology of instanton moduli space 
\cite{Belavin:2011pp, Nishioka:2011jk, Belavin:2011sw} 
(see also \cite{Bonelli:2012ny}). 
Here $\mathcal{H}$ is the affine Heisenberg algebra, and 
$p$, which parametrizes the central charge in the 2D CFT, 
is related to the ratio $\epsilon_1/\epsilon_2$ 
of the $\Omega$-deformation parameters 
$\epsilon_1$, $\epsilon_2$ on ${\IC}^2/{\IZ}_n$ 
(see eqns.~\eqref{omega_pp} and \eqref{cc_paraW}). 
The 2D CFT, in particular, has the $\mathcal{W}^{\, para}_{N, n}$ 
($n$-th parafermion $\mathcal{W}_N$) symmetry 
\cite{Fateev:1985mm, Gepner:1987sm} described by 
the third (coset) factor \cite{Bais:1987zk, Christe:1988vc, Bowcock:1988vs} 
in the algebra $\mathcal{A}(N,n;p)$. 
When $n=1$, it gives the $\mathcal{W}_N$ algebra in 
\cite{Zamolodchikov:1985wn, Fateev:1987vh, Fateev:1987zh} 
which contains higher spin currents.

For the gauge theory with an adjoint hypermultiplet, 
the AGT-corresponding CFT lives on a torus $T^2$. 
In the case of $\epsilon_1+\epsilon_2=0$ (corresponding to $p \to \infty$), 
the $U(N)$ instanton partition function on ${\IC}^2/{\IZ}_n$ 
for the massless adjoint hypermultiplet 
yields the partition function of an $\mathcal{N}=4$ twisted Yang-Mills theory 
enumerating torus fixed points on the moduli space of 
instantons, which is labelled by $N$-tuples of $n$-coloured Young diagrams 
$(Y_{1}^{\sigma_{1}},\ldots,Y_{N}^{\sigma_{N}})$ with 
${\IZ}_n$ charges $\sigma_I \in \{0,1,\ldots,n-1\}$, $1\le I \le N$ 
(see \textit{e.g.} \cite{Pestun:2007rz, Okuda:2010ke}). 
The twisted partition function is well-known to give 
a character of the 2D CFT 
\cite{Nakajima:1994nid, Nakajima:1998, Vafa:1994tf}, where 
a string theory interpretation is given in \cite{Dijkgraaf:2007sw}.

\subsection{AGT correspondence for minimal models}

The AGT correspondence for minimal models was proposed in \cite{Bershtein:2014qma, Alkalaev:2014sma, Belavin:2015ria} 
(see also \cite{Santachiara:2010bt, Estienne:2011qk} for early works) when $n=1$, 
and it was generalized to $n \ge 2$ in \cite{Foda:2019msm}. 
When $p$ in the algebra $\mathcal{A}(N,n;p)$ is an integer with $p \ge N$, 
one finds that the $U(N)$ instanton partition function on ${\IC}^2/{\IZ}_n$ has 
\textit{non-physical poles} which need to be removed and 
supposed to correspond to 
$\mathcal{W}^{\, para}_{N, n}$ minimal model null states.
The poles are parametrized by positive integers 
$r_I$ and $s_I$, $0 \le I < N$, with 
$\sum_{I=0}^{N-1}r_I = p$ and $\sum_{I=0}^{N-1}s_I = p+n$, and 
shown to be removed by imposing \textit{Burge conditions}
$$
Y_{I,i}^{\sigma_I} \ge Y_{I+1, i+r_I-1}^{\sigma_{I+1}} - s_I+1 \ \
\textrm{for}\ i \ge 1,\ 0 \le I < N,
$$
on $N$-tuples of $n$-coloured Young diagrams  $(Y_{1}^{\sigma_{1}},\ldots,Y_{N}^{\sigma_{N}})$, 
where $Y_{0}^{\sigma_{0}}=Y_N^{\sigma_N}$, and 
the ${\IZ}_n$ charges $\sigma_I$ satisfy 
the ${\IZ}_n$ \textit{charge conditions} 
$\sigma_I - \sigma_{I+1} \equiv - r_I + s_I$ $(\mathrm{mod}\ n)$, 
$0 \le I < N$, with $\sigma_0=\sigma_N$.

Following the algebra $\mathcal{A}(N,n;p)$, 
the generating functions of the coloured Young diagrams with 
the Burge conditions and the ${\IZ}_n$ charge conditions, 
that we will refer as \textit{Burge-reduced generating functions}, 
are expected to be decomposed into $\widehat{\mathfrak{sl}}(n)_N$ 
WZW (Wess-Zumino-Witten model) characters \cite{kac.book.1990} and 
$\mathcal{W}^{\, para}_{N, n}$ $(p,p+n)$-minimal model characters  
(branching functions of the coset factor in $\mathcal{A}(N,n;p)$ 
\cite{Christe:1988vc, Bouwknegt:1990fb}) 
up to a Heisenberg factor. 
In \cite{Foda:2019msm} we discussed the special case $p=N$ in which 
the coset factor in $\mathcal{A}(N,n;p)$ is trivialized and, 
using the results in the crystal graph theory of \cite{DJKMO:1989}, 
showed that the Burge-reduced generating functions indeed give 
the $\widehat{\mathfrak{sl}}(n)_N$ WZW characters. 
The aim of this paper is to generalize it to integral $p \ge N$ and propose how 
the $\mathcal{W}^{\, para}_{N, n}$ $(p,p+n)$-minimal model characters 
are obtained from the Burge-reduced generating functions.

\subsection{Plan of the paper}

In Section \ref{sec:agt_instanton}, we summarize 
the minimal ingredients about the AGT correspondence for minimal models 
and introduce $SU(N)$ Burge-reduced generating functions of 
$n$-coloured Young diagrams by subtracting the overall $U(1)$ factor 
corresponding to $\mathcal{H}$. 
We then recall that the Burge-reduced generating functions 
in the special case $p=N$ agree with 
the $\widehat{\mathfrak{sl}}(n)_N$ WZW characters. 
In Section \ref{sec:w_minimal_instanton} we generalize it to $p \ge N$ and 
propose Conjecture \ref{conj:suN_red_ch} which states 
a decomposition of the Burge-reduced generating functions 
into the $\widehat{\mathfrak{sl}}(n)_N$ WZW characters and 
the $\mathcal{W}^{\, para}_{N, n}$ $(p,p+n)$-minimal model characters. 
The conjectural decomposition formula is considered to be a generalization of 
a character decomposition formula in \cite{JimboMiwa:85, Hasegawa:89} 
for $p \to \infty$ established in the context of the level-rank duality 
\cite{Frenkel:1982, Naculich:1990hg, Nakanishi:1990hj}. 
We check the conjecture, by extracting 
the $\mathcal{W}^{\, para}_{N, n}$ $(p,p+n)$-minimal model characters 
from the Burge-reduced generating functions, 
for $(N,n,p)=(2,2,4), (3,3,4)$ in Section \ref{sec:examples} 
and for $(N,n,p)=(2,3,3), (2,4,4), (3,2,4), (4,2,5)$ 
in Appendix \ref{app:add_examples}. 
Section \ref{sec:summary} is devoted to summary and outlook. 
In Appendix \ref{app:string_fn} we summarize some string functions, 
and in Appendix \ref{app:hw_ex} we give some examples of 
the dominant integral weights of $\widehat{\mathfrak{sl}}(n)_N$ 
which are dual to the dominant integral weights of $\widehat{\mathfrak{sl}}(N)_n$ 
defined in Section \ref{subsec:dual_dw}.

\subsection{Notation}

We use the following notation of affine Lie algebras 
(see \cite[Appendix A]{Foda:2019msm}). 

\begin{figure}[t]
\centering
\includegraphics[width=80mm]{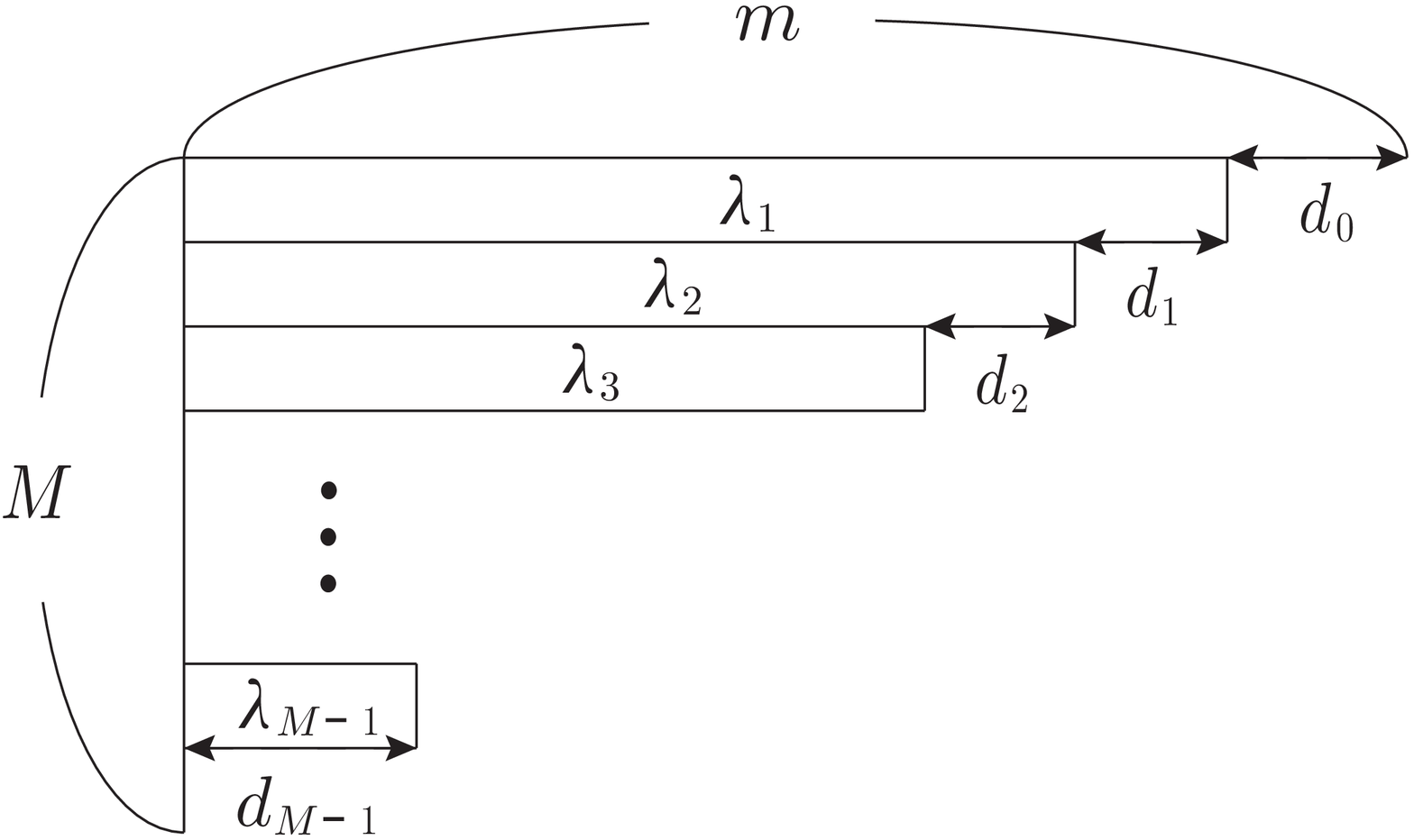}
\caption{The partition $\partit(\Lambda)$ for 
a dominant weight $\Lambda=[d_0,d_1,\ldots,d_{M-1}] \in P^{+}_{\dimaff,\levelaff}$.}
\label{fig:partition_dw}
\end{figure}

Consider the affine Lie algebra $\slchap{\dimaff}$, and 
define the index sets $\indaff=\{0,1,\ldots,\dimaff-1\}$ and 
$\indfin=\{1,2,\ldots,\dimaff-1\}$. 
Let $\alpha_i$ and $\Lambda_i$ for $i \in \indaff$ be the simple roots and 
fundamental weights of $\slchap{\dimaff}$. For the standard inner product 
$\inner{\cdot}{\cdot}$ they satisfy
\begin{align}
\inner{\alpha_i}{\alpha_j}=A_{ij},
\quad
\inner{\alpha_i}{\Lambda_j}=\delta_{ij},
\quad
\inner{\Lambda_i}{\Lambda_j}=\mathrm{min}\{i, j\}-\frac{i\, j}{\dimaff}
\end{align}
for $i,j \in \indaff$, where $A$ is the Cartan matrix of 
$\slchap{\dimaff}$. For the inner product we use a notation 
$|\Lambda|^2=\inner{\Lambda}{\Lambda}$. 
The Weyl vector $\rho$ is defined by $\rho=\sum_{i \in \indaff} \Lambda_i$. 
The level-$\levelaff$ weight lattice $P_{\dimaff,\levelaff}$, 
the level-$\levelaff$ dominant weight lattice $P^{+}_{\dimaff,\levelaff}$, 
the level-$\levelaff$ regular dominant weight lattice 
$P^{++}_{\dimaff,\levelaff}$ and 
the root lattice $\overline{Q}_{\dimaff}$ are defined by
\begin{align}
\begin{split}
P_{\dimaff,\levelaff}&=
\left\{\, \Lambda \in \bigoplus_{i \in \indaff} {\IZ}\, \Lambda_i\, 
\bigg|\, \Lambda=\sum_{i \in \indaff} d_i\, \Lambda_i,\ \
\sum_{i \in \indaff} d_i=\levelaff\, \right\}\,,
\\
P^{+}_{\dimaff,\levelaff}&=
P_{\dimaff,\levelaff}\cap
\bigoplus_{i \in \indaff} {\IZ}_{\ge 0}\, \Lambda_i\,,
\qquad
P^{++}_{\dimaff,\levelaff}=
P_{\dimaff,\levelaff}\cap
\bigoplus_{i \in \indaff} {\IZ}_{> 0}\, \Lambda_i\,,
\\
\overline{Q}_{\dimaff}&=
\bigoplus_{i \in \indfin} {\IZ}\, \alpha_i\,.
\label{lattices}
\end{split}
\end{align}
We often use the notation $[d_0,d_1,\ldots,d_{M-1}]$ 
of Dynkin labels to denote 
$\Lambda=\sum_{i \in \indaff} d_i\, \Lambda_i$. 
A partition $\lambda=(\lambda_1, \lambda_2, \ldots)$ 
for $\Lambda=[d_0,d_1,\ldots,d_{M-1}] \in P^{+}_{\dimaff,\levelaff}$ 
is introduced by
\begin{align}
\lambda_i=
\begin{cases}
\sum_{j=i}^{M-1}d_j
&\textrm{if}\ \ 1 \le i < M,
\\
0
&\textrm{if}\ \ i \ge M,
\end{cases}
\label{partition_L}
\end{align}
and denoted by $\partit(\Lambda)$ (see Figure \ref{fig:partition_dw}). 
The transposed partition of $\partit(\Lambda)$ is denoted by 
$\partit(\Lambda)^T=(\lambda_1^T, \lambda_2^T, \ldots)$, and 
one can write $\Lambda=\partit^{-1}(\lambda)$ when $m$ is specified.

\section{AGT correspondence for $U(N)$ instanton counting on 
${\IC}^2/{\IZ}_n$}\label{sec:agt_instanton}

\textit{\noindent
In this section, we recall some contents 
in Sections 2, 3, 4 and 5 of \cite{Foda:2019msm} about 
the $U(N)$ instanton counting on ${\IC}^2/{\IZ}_n$, 
the AGT correspondence for minimal models and the Burge conditions.}

\subsection{$U(N)$ instanton counting on 
${\IC}^2/{\IZ}_n$}\label{subsec:gauge}

The instanton moduli space $\mathcal{M}_{N, n}$ of $U(N)$ instantons 
on ${\IC}^2/{\IZ}_n$ is 
characterized by the fixed point set of $U(1)^2 \times U(1)^{N}$ 
torus action on $\mathcal{M}_{N, n}$, where the $U(1)^2$ torus is 
generated by the $\Omega$-deformation parameters $\epsilon_1, \epsilon_2$, 
through 
$(z_1, z_2)\in {\IC}^2\ \to\  (\e^{\epsilon_1} z_1, \e^{\epsilon_2} z_2)$, 
and the $U(1)^{N}$ torus is generated by the Coulomb parameters 
$a_{I}$, $I=1, 2,\ldots, N$, which parametrize the Cartan subalgebra of $U(N)$. 
The fixed point set has the colour coding 
induced by the ${\IZ}_n$ orbifold of ${\IC}^2$ as 
$(z_1, z_2)\ \to\  (\e^{\frac{2 \pi i}{n}\, \sigma} z_1, \e^{-\frac{2 \pi i}{n}\, \sigma} z_2)$, $\sigma=0,1,\ldots,n-1$, 
and is described by $N$-tuples of $n$-coloured Young diagrams 
$\bY^{\bsig}=(Y_{1}^{\sigma_{1}},\ldots,Y_{N}^{\sigma_{N}})$ as follows 
\cite{KronheimerNakajima90, Fucito:2004ry}.

A coloured Young diagram $Y^{\sigma}$, with ${\IZ}_n$ charge 
$\sigma \in \{0,1,\ldots,n-1\}$, 
is a Young diagram whose box at position 
$(i,j)\in Y^{\sigma}$ has a colour $\sigma-i+j \ (\mathrm{mod}\ n)$. 
The length of the $i$-row in $Y^{\sigma}$ is denoted by $Y_i^{\sigma}$, 
and the total number of boxes in $Y^{\sigma}$ is 
$|Y^{\sigma}|=\sum_{i}Y_i^{\sigma}$.

Let $k_i$, $0 \le i < n$, be the total number of boxes with colour $i$ in 
$\bY^{\bsig}$, and $\mathcal{P}_{\bsig;\bdel \bk}$ be 
the set of $N$-tuples of $n$-coloured Young diagrams $\bY^{\bsig}$ labelled by 
the ${\IZ}_n$ charges $\bsig=(\sigma_1,\ldots,\sigma_N)$ and 
$\bdel \bk=(\delta k_1,\ldots, \delta k_{n-1})$, where 
$\delta k_i=k_i-k_0$. 
The charges $\bsig$ define the non-negative integers $N_i$ as 
the number of coloured Young diagrams with charge $i$, and 
we have
\begin{align}
\left|\bY^{\bsig}\right|:=\sum_{I=1}^N \left|Y_I^{\sigma_I}\right|
=\sum_{i=0}^{n-1} k_i,\qquad
N=\sum_{i=0}^{n-1} N_i.
\label{young_num}
\end{align}
As a characterization of the $U(N)$ instantons on ${\IC}^2/{\IZ}_n$, 
consider the first Chern class 
$c_1 = \sum_{i=0}^{n-1} \fc_{i}\, c_1(\mathcal{T}_{i})$ of the gauge bundle. 
Here $c_1(\mathcal{T}_{i})$ is the first Chern class of 
an individual vector bundle $\mathcal{T}_{i}$ 
associated with the ${\IZ}_n$ orbifold, 
where $c_1(\mathcal{T}_{0})=0$, and 
\begin{align}
\fc_{i} = 
N_{i} + \delta k_{i-1} - 2\delta k_{i} + \delta k_{i+1} = 
N_{i} - \sum_{j=0}^{n-1} A_{ij}\, \delta k_j,
\label{inst_chern}
\end{align}
for $0 \le i < n$, 
where $k_n=k_0$, $k_{-1}=k_{n-1}$, and 
$A$ denotes the Cartan matrix of $\widehat{\mathfrak{sl}}(n)$.

Now, it is useful to identify the non-negative integers $N_i$, 
$0 \le i < n$, with the Dynkin labels of $\widehat{\mathfrak{sl}}(n)$ 
in the level-$N$ dominant weight lattice as 
$\bN=[N_0,N_1,\ldots,N_{n-1}]\in P^{+}_{n,N}$. 
We then introduce a generating function which enumerates 
the fixed points of $U(1)^2 \times U(1)^{N}$ torus action on 
the instanton moduli space $\mathcal{M}_{N, n}$. 

\begin{defi}\label{def:t_ref_ch}
For $\bN=[N_0,N_1,\ldots,N_{n-1}] \in P^{+}_{n,N}$, 
the $SU(N)$ $\ft$-refined generating function of $n$-coloured Young diagrams 
is defined by
\begin{align}
\widehat{X}_{\bN}(\fq, \bft)=
\sum_{\bdel \bk \in {\IZ}^{n-1}}
\widehat{X}_{\bsig;\bdel \bk}(\fq)\, 
\prod_{i=1}^{n-1} \ft_{i}^{\, \fc_i(\bdel \bk)},
\label{inst_ch_gen}
\end{align}
where $\fc_i(\bdel \bk)=N_{i} + \delta k_{i-1} - 2\delta k_{i} + \delta k_{i+1}$ 
are the Chern classes \eqref{inst_chern}, and 
\begin{align}
\widehat{X}_{\bsig;\bdel \bk}(\fq)=
\left(\fq;\fq \right)_{\infty}\, 
\sum_{\bY^{\bsig}\in \mathcal{P}_{\bsig;\bdel \bk}}
\fq^{\, \frac{1}{n}\, \left|\bY^{\bsig}\right|}\,.
\label{inst_ch}
\end{align}
Here $\widehat{X}_{\bsig;\bdel \bk}(\fq)$ does not depend on 
the ordering of the ${\IZ}_n$ charges $\bsig$ and 
\eqref{inst_ch_gen} is well-defined,%
\footnote{\,
If the ordering $\sigma_1 \ge \sigma_2 \ge \ldots \ge \sigma_N$ is assumed, 
the ${\IZ}_n$ charges are described by the partition \eqref{partition_L} as 
$(\sigma_1,\sigma_2,\ldots,\sigma_N,0,0,\ldots)=\partit(\bN)^T$.
}
and the prefactor 
$\left(\fq;\fq \right)_{\infty}=\prod_{n=1}^{\infty}(1-\fq^n)$ 
subtracts the $U(1)$ factor in $U(N)$ gauge theory.
\end{defi}

As mentioned in the introduction, 
the generating function \eqref{inst_ch} originates with 
the $U(N)$ instanton partition function 
on ${\IC}^2/{\IZ}_n$ with a massless adjoint hypermultiplet 
in the case of $\epsilon_1+\epsilon_2=0$ 
and pertains to the partition function of 
an $\mathcal{N}=4$ twisted Yang-Mills theory on ${\IC}^2/{\IZ}_n$ 
\cite{Vafa:1994tf} (see \cite{Dijkgraaf:2007sw} 
for a string theory interpretation).

\subsection{Algebra $\mathcal{A}(N,n;p)$}

For a $U(N)$ gauge theory on ${\mathbb {C}}^2/{\mathbb {Z}}_n$ with $\Omega$-deformation, the relevant AGT-corresponding CFT 
possesses symmetry algebra
\begin{align}
\mathcal{A}(N,n;p)=\mathcal{H} \oplus 
\widehat{\mathfrak{sl}}(n)_N \oplus 
\frac{\widehat{\mathfrak{sl}}(N)_n \oplus 
\widehat{\mathfrak{sl}}(N)_{p-N}}{\widehat{\mathfrak{sl}}(N)_{n+p-N}},
\label{alg_A}
\end{align}
which acts on the equivariant cohomology of $\mathcal{M}_{N, n}$
\cite{Belavin:2011pp, Nishioka:2011jk, Belavin:2011sw} 
(see also \cite{Nakajima:1994nid, Nakajima:1998} for the early notable works 
by Nakajima), 
where $\mathcal{H}$ is the affine Heisenberg algebra. 
This implies that the AGT-corresponding CFT is a combined system of 
$\widehat{\mathfrak{sl}}(n)_N$ WZW model with the additional $\mathcal{H}$ 
symmetry and a 2D CFT with 
the $\mathcal{W}^{\, para}_{N, n}$ ($n$-th parafermion $\mathcal{W}_N$) symmetry described by the coset \cite{Bais:1987zk, Christe:1988vc, Bowcock:1988vs}
\begin{align}
\frac{\widehat{\mathfrak{sl}}(N)_n \oplus 
\widehat{\mathfrak{sl}}(N)_{p-N}}{\widehat{\mathfrak{sl}}(N)_{n+p-N}}.
\label{coset_A}
\end{align}
The parameter $p$ is related to 
the $\Omega$-deformation parameters $\epsilon_1$, $\epsilon_2$ by
\begin{align}
\frac{\epsilon_1}{\epsilon_2} = -1-\frac{n}{p},
\label{omega_pp}
\end{align}
and controls the central charge of the 2D CFT with $\mathcal{W}^{\, para}_{N, n}$ symmetry by 
\begin{align}
c\left(\mathcal{W}^{\, para}_{N, n}\right)=
\frac{n\, (N^2-1)}{N+n} 
\ll 1 - \frac{N\, (N+n)}{p\,(p+n)} \rr.
\label{cc_paraW}
\end{align}
Here, if we take the limit $p\to \infty$ corresponding to 
$\epsilon_1+\epsilon_2=0$, the algebra $\mathcal{A}(N,n;p)$ is formally 
reduced to $\mathcal{H} \oplus 
\widehat{\mathfrak{sl}}(n)_N \oplus \widehat{\mathfrak{sl}}(N)_n$. 
Since the central charge of $\widehat{\mathfrak{sl}}(n)_N$ WZW model is
\begin{align}
c(\widehat{\mathfrak{sl}}(n)_N)=\frac{N(n^2-1)}{N+n},
\end{align} 
the AGT-corresponding CFT with this symmetry algebra has the central charge
\begin{align}
1+\frac{N(n^2-1)}{N+n}+\frac{n(N^2-1)}{N+n} = Nn,
\label{cc_p_inf}
\end{align}
and is considered to be described by $Nn$ free fermions 
(see below \eqref{conf_embed} and \textit{e.g.} \cite{Dijkgraaf:2007sw}).

\subsection{Burge conditions}

When
\begin{align}
p \in {\IN}\ \ \textrm{with}\ \ p\ge N,
\end{align}
the ratio of the $\Omega$-deformation parameters \eqref{omega_pp} 
becomes rational, and then 
the instanton partition function in 4D $\mathcal{N}=2$ $U(N)$ Yang-Mills theory 
on ${\mathbb {C}}^2/{\mathbb {Z}}_n$ has non-physical poles \cite{Foda:2019msm} 
(see also \cite{Bershtein:2014qma, Alkalaev:2014sma, Belavin:2015ria} 
for early works in the case of $n=1$). 
By the AGT correspondence, these poles should correspond to 
the null states in $\mathcal{W}^{\, para}_{N, n}$ $(p,p+n)$-minimal models, 
which are described by the coset \eqref{coset_A}, and 
are parametrized by positive integers 
$r_I$ and $s_I$, $0 \le I < N$, with
\begin{align}
\sum_{I=0}^{N-1}r_I = p,\qquad 
\sum_{I=0}^{N-1}s_I = p+n.
\label{sr_0}
\end{align}
Similarly to $\bN=[N_0,N_1,\ldots,N_{n-1}]\in P^{+}_{n,N}$, 
in what follows, we identify the positive integers 
$r_I$ and $s_I$, $0 \le I < N$, with 
the Dynkin labels of $\widehat{\mathfrak{sl}}(N)$ 
in the level-$n$ regular dominant weight lattices as 
$\br=[r_0,r_1,\ldots,r_{N-1}]\in P^{++}_{N,p}$ 
and 
$\bs=[s_0,s_1,\ldots,s_{N-1}]\in P^{++}_{N,p+n}$.

One finds that the poles can be removed by imposing the Burge conditions 
\cite{Bershtein:2014qma, Alkalaev:2014sma, Belavin:2015ria, Foda:2019msm} 
(see also \cite{Burge, Foda:1997du, GesselKrattenthaler, Feigin:2010qea1, Feigin:2010qea2} for Burge conditions),
\begin{align}
Y_{I,i}^{\sigma_I} \ge Y_{I+1, i+r_I-1}^{\sigma_{I+1}} - s_I+1 \ \
\textrm{for}\ i \ge 1,\ 0 \le I < N,
\label{burge_cd}
\end{align}
on $N$-tuples of $n$-coloured Young diagrams 
$\bY^{\bsig}=(Y_1^{\sigma_1},\ldots,Y_N^{\sigma_N})$, 
where $Y_{0}^{\sigma_{0}}=Y_N^{\sigma_N}$. 
The ${\IZ}_n$ charges $\sigma_I$ are related to 
$\br$ and $\bs$ by the ${\IZ}_n$ charge conditions \cite{Foda:2019msm}
\begin{align}
\sigma_I - \sigma_{I+1} \equiv - r_I + s_I \quad (\mathrm{mod}\ n),
\quad
0 \le I < N,
\label{charge_condition}
\end{align}
where we set $\sigma_0=\sigma_N$.

\subsection{Burge-reduced generating functions}

Let $\mathcal{C}^{\br,\bs}_{\bsig;\bdel \bk}$ be the subset of 
$\mathcal{P}_{\bsig;\bdel \bk}$,
\begin{align}
\mathcal{C}^{\br,\bs}_{\bsig;\bdel \bk} \subset 
\mathcal{P}_{\bsig;\bdel \bk},
\end{align}
whose elements satisfy the Burge conditions \eqref{burge_cd} and 
the ${\IZ}_n$ charge conditions \eqref{charge_condition}. 
We now introduce Burge-reduced generating functions of 
coloured Young diagrams by subtracting the overall $U(1)$ factor 
corresponding to $\mathcal{H}$. 

\begin{defi}\label{def:t_ref_red_ch}
For $\bN=[N_0,N_1,\ldots,N_{n-1}] \in P^{+}_{n,N}$, 
the $SU(N)$ $\ft$-refined Burge-reduced generating function of 
$n$-coloured Young diagrams, 
which is reduced by the Burge conditions \eqref{burge_cd} 
for $\br=[r_0,r_1,\ldots,r_{N-1}] \in P^{++}_{N,p}$ and 
$\bs=[s_0,s_1,\ldots,s_{N-1}] \in P^{++}_{N,p+n}$, is defined by
\begin{align}
\widehat{X}_{\bN}^{\, \br,\bs}(\fq, \bft)=
\sum_{\bdel \bk \in {\IZ}^{n-1}}
\widehat{X}_{\bsig;\bdel \bk}^{\, \br,\bs}(\fq)\, 
\prod_{i=1}^{n-1} \ft_{i}^{\, \fc_i(\bdel \bk)},
\label{rs_red_ch_gen}
\end{align}
where $\fc_i(\bdel \bk)=N_i+\delta k_{i-1}-2\delta k_{i}+\delta k_{i+1}$ 
are the Chern classes \eqref{inst_chern}, and 
\begin{align}
\widehat{X}_{\bsig;\bdel \bk}^{\, \br,\bs}(\fq)=
\left(\fq;\fq\right)_{\infty}\,
\sum_{\bY^{\bsig}\in \mathcal{C}^{\br,\bs}_{\bsig;\bdel \bk}}
\fq^{\, \frac{1}{n}\, \left|\bY^{\bsig}\right|}\,.
\label{rs_red_ch}
\end{align}
Here, for fixed $\br$ and $\bs$ the ${\IZ}_n$ charge conditions \eqref{charge_condition} fix the charges $\bsig$ up to 
the shifts $\sigma_I \to \sigma_I - k$ modulo $n$ by $k \in {\IZ}_n$ and  
the cyclic permutations $\sigma_I \to \sigma_{I-\theta}$ by 
$\theta \in {\IZ}_N$, where $\sigma_{I+N}=\sigma_I$ and 
the latter ambiguities exist only if 
$s_0-r_0 \equiv s_1-r_1 \equiv \ldots \equiv s_{N-1}-r_{N-1}$ 
$(\mathrm{mod}\ n)$. 
Once we fix $\bN$, the former ambiguities are fixed. 
As seen from the Burge conditions \eqref{burge_cd}, 
$\widehat{X}_{\bsig;\bdel \bk}^{\, \br,\bs}(\fq)$ 
is invariant under the cyclic permutations $\sigma_I \to \sigma_{I-\theta}$, 
$r_I \to r_{I-\theta}$ and $s_I \to s_{I-\theta}$, 
where $r_{I+N}=r_I$ and $s_{I+N}=s_I$, 
and so \eqref{rs_red_ch_gen} is well-defined. 
This also implies
\begin{align}
\widehat{X}_{\bN}^{\, \br,\bs}(\fq, \bft)=
\widehat{X}_{\bN}^{\, \br^{(\theta)},\bs^{(\theta)}}(\fq, \bft),
\quad
\theta \in {\IZ}_N,
\label{inv_b_red_g}
\end{align}
where $\br^{(\theta)}=[r^{(\theta)}_0, \ldots, r^{(\theta)}_{N-1}]$ 
with $r^{(\theta)}_{I}=r_{I-\theta}$ and 
$\bs^{(\theta)}=[s^{(\theta)}_0, \ldots, s^{(\theta)}_{N-1}]$ 
with $s^{(\theta)}_{I}=s_{I-\theta}$.
\end{defi}

Consider the special case $p=N$ in which the algebra $\mathcal{A}(N,n;p)$ is 
reduced to $\mathcal{A}(N,n;N)=\mathcal{H} \oplus 
\widehat{\mathfrak{sl}}(n)_N$, and then 
$\br=\boldsymbol{1}=\rho$ is fixed by \eqref{sr_0}. 
In \cite[Corollary 5.5]{Foda:2019msm}, 
using the results of \cite{DJKMO:1989}, it was shown that 
the $\ft$-refined Burge-reduced generating function \eqref{rs_red_ch_gen} 
for $\bN=[N_0,N_1,\ldots,N_{n-1}]\in P^{+}_{n,N}$ 
agrees with the $\widehat{\mathfrak{sl}}(n)_N$ WZW character as
\footnote{\, 
When $\bN$ is fixed, by the ${\IZ}_n$ charge conditions 
$\sigma_I - \sigma_{I+1} \equiv s_I - 1$ $(\mathrm{mod}\ n)$ 
with $\sum_{I=0}^{N-1}(s_I-1)=n$ and $s_I-1\ge 0$, 
the generating function 
$\widehat{X}_{\bN}^{\, \boldsymbol{1},\bs}(\fq, \bft)$ is 
ambiguous only for the cyclic permutations 
$s_I \to s_{I-\theta}$ by $\theta \in {\IZ}_N$. 
By \eqref{inv_b_red_g}, this is not the actual ambiguity of 
$\widehat{X}_{\bN}^{\, \boldsymbol{1},\bs}(\fq, \bft)$, 
and one can assume 
$\sigma_1 \ge \sigma_2 \ge \ldots \ge \sigma_N$ and  
$s_I=\sigma_I-\sigma_{I+1}+1+n\,\delta_{I,0}$.
\label{fn:sp_wzw_ch}}
\begin{align}
\widehat{X}_{\bN}^{\, \boldsymbol{1},\bs}(\fq, \bft)=
\fq^{\, w_{\bN} - h_{\bN}}\,
\chi_{\bN}^{\widehat{\mathfrak{sl}}(n)_N}(\fq, \hat{\bft})
\quad
\textit{if}\ \ p=N,
\label{wzw_ch_comb_c}
\end{align}
where
\begin{align}
\hat{\ft}_{i}=\fq^{- \frac{i\, (n-i)}{2n}}\, \ft_i,
\qquad
w_{\bN}=\frac{\inner{\bN}{\rho}}{n}
=\sum_{i=1}^{n-1}\frac{i\, (n-i)}{2n}\, N_i,
\qquad
h_{\bN}=\frac{\inner{\bN}{\bN + 2\rho}}{2(n + N)}.
\label{normal_fc_w}
\end{align}
Here the $\widehat{\mathfrak{sl}}(n)_N$ WZW character is defined by 
(an overall normalization factor 
$\fq^{\,- \frac{1}{24}\, c(\widehat{\mathfrak{sl}}(n)_N)}$ 
is further introduced in the literature),
\begin{align}
\chi_{\bN}^{\widehat{\mathfrak{sl}}(n)_N}(\fq, \hat{\bft}) =
\mathrm{Tr}_{L(\bN)}\, 
\fq^{\, L_0}\, 
\prod_{i=1}^{n-1} \hat{\ft}_{i}^{\, H_i},
\label{wzw_ch}
\end{align}
where $L(\bN)$ is the level-$N$ irreducible highest-weight module of 
$\widehat{\mathfrak{sl}}(n)$, and 
the Virasoro generator $L_0$ and 
the Chevalley elements $H_i$ in the Cartan subalgebra 
of $\widehat{\mathfrak{sl}}(n)$ act on 
the modules in the representation of a highest-weight state 
with the eigenvalues $h_{\bN}$ and $N_i$, respectively. 
As in \eqref{rs_red_ch_gen}, we now expand 
the $\widehat{\mathfrak{sl}}(n)_N$ WZW character \eqref{wzw_ch} as
\begin{align}
\chi_{\bN}^{\widehat{\mathfrak{sl}}(n)_N}( \fq, \hat{\bft}) = 
\fq^{\, h_{\bN}}
\sum_{\bdel \bk \in {\IZ}^{n-1}}
a^{\bN}_{\bfc(\bdel \bk)}(\fq)\, 
\prod_{i=1}^{n-1} \hat{\ft}_{i}^{\, \fc_i(\bdel \bk)}=
\sum_{\bdel \bk \in {\IZ}^{n-1}}
\hat{a}^{\bN}_{\bfc(\bdel \bk)}(\fq)\, 
\prod_{i=1}^{n-1} \ft_{i}^{\, \fc_i(\bdel \bk)},
\label{wzw_string}
\end{align}
where $a^{\bN}_{\bfc}(\fq)$ is known as 
a (normalized) $\widehat{\mathfrak{sl}}(n)$ string function of level-$N$ 
(see also \eqref{gen_string_fn_Nn}) and 
\begin{align}
\hat{a}^{\bN}_{\bfc}(\fq)=\fq^{\, h_{\bN} - w_{\bfc}}\,
a^{\bN}_{\bfc}(\fq),
\label{hata_string_n}
\end{align}
is also introduced, where $\bfc=\bfc(\bdel \bk)$. 
From \eqref{wzw_ch_comb_c}, by comparing \eqref{rs_red_ch_gen} 
with \eqref{wzw_string} one obtains \cite{Foda:2019msm}
\begin{align}
\widehat{X}_{\bsig;\bdel \bk}^{\, \boldsymbol{1},\bs}(\fq)=
\fq^{\, \frac{1}{n}\sum_{i=1}^{n-1}\delta k_i}\,
a^{\bN}_{\bfc(\bdel \bk)}(\fq)\, ,
\label{instanton_string}
\end{align}
where $w_{\bN}-w_{\bfc(\bdel \bk)}=\frac{1}{n}\sum_{i=1}^{n-1}\delta k_i$ 
was used.

\section{$\mathcal{W}^{\, para}_{N, n}$ minimal model characters 
from the instanton counting}\label{sec:w_minimal_instanton}

\textit{\noindent
In this section, we first recall a formula of 
$\mathcal{W}^{\, para}_{N, n}$ $(p,p+n)$-minimal model characters 
(branching functions), and then propose Conjecture \ref{conj:suN_red_ch} about 
how the $\ft$-refined Burge-reduced generating functions of 
coloured Young diagrams 
are decomposed into the characters following the algebra $\mathcal{A}(N,n;p)$.
}

\subsection{$\mathcal{W}^{\, para}_{N, n}$ minimal model characters}

We introduce a normalized $\widehat{\mathfrak{sl}}(N)$ string function 
$\hat{c}^{\, \bel}_{\,\bm}(\fq)$ of level-$n$ for a dominant highest-weight 
$\bel=[\ell_0,\ell_1,\ldots,\ell_{N-1}]\in P^{+}_{N,n}$ and 
a maximal-weight $\bm=[m_0,m_1,\ldots,m_{N-1}]\in P_{N,n}$ 
by normalizing the $\widehat{\mathfrak{sl}}(N)$ string function 
$a^{\, \bel}_{\bm}(\fq)$ of level-$n$ in \eqref{wzw_string} 
with the exchange $N \leftrightarrow n$ as
\footnote{\,
A maximal-weight $\bm$ in $P_{N,n}$ is obtained from a dominant maximal-weight 
in $P^+_{N,n}$ by 
an action of the affine Weyl group of $\widehat{\mathfrak{sl}}(N)$, and 
the string function is invariant under the action 
(see Proposition 2.12 (a) and eqn.~(2.17) in \cite{Kac:1984mq}). 
\label{fn:maximal_wt}}
\begin{align}
\hat{c}^{\, \bel}_{\,\bm}(\fq)=
\fq^{\, h_{\bel} - \frac{1}{2n}|\bm|^2}\,
a^{\, \bel}_{\bm}(\fq).
\label{gen_string_fn_Nn}
\end{align}
Here $\hat{c}^{\, \bel}_{\,\bm}(\fq)$ is related to 
the string function $c^{\, \bel}_{\,\bm}(\fq)$ in 
\cite{Kac:1984mq, kac.book.1990} by 
$c^{\, \bel}_{\,\bm}(\fq)=
\fq^{\, -\frac{1}{24}\, c(\widehat{\mathfrak{sl}}(N)_n)}\, 
\hat{c}^{\, \bel}_{\,\bm}(\fq)$, where 
$c(\widehat{\mathfrak{sl}}(N)_n)=\frac{n(N^2-1)}{n+N}$ is 
the central charge of $\widehat{\mathfrak{sl}}(N)_n$ WZW model. 
Note that, for non-zero string functions, 
the highest-weight $\bel$ and the maximal-weight $\bm$ 
should satisfy
\begin{align}
\sum_{I=1}^{N-1} \left(\ell_I - m_I\right)\Lambda_I \in 
\overline{Q}_{N},
\quad i.e. \quad
\sum_{I=1}^{N-1} I \, (\ell_I - m_I) \equiv 0\ \
(\mathrm{mod}\ N),
\label{string_condition}
\end{align}
where $\overline{Q}_{N}$ is the root lattice in \eqref{lattices}. 
Note also that, under the outer automorphisms of $\widehat{\mathfrak{sl}}(N)$ 
which cyclically permutes the Dynkin labels as 
$\ell_I \to \ell_{I-\theta}$ and $m_I \to m_{I-\theta}$ 
for all $I=0,1,\ldots,N-1$ by 
$\theta \in {\IZ}_N$,
the string functions \eqref{gen_string_fn_Nn} are invariant, 
where we set $\ell_{I+N}=\ell_I$ and $m_{I+N}=m_I$. 
Here, by 
$\hat{a}^{\, \bel}_{\bm}(\fq)=
\fq^{\,\frac{1}{2n}|\bm|^2-w_{\bm}}\,\hat{c}^{\, \bel}_{\,\bm}(\fq)$, 
and 
$\frac{1}{2n}|\bm|^2-w_{\bm}=\frac{1}{2nN}\sum_{0\le I<J \le N-1} 
(I-J)(N+I-J)m_Im_J$, 
the normalized string functions $\hat{a}^{\, \bel}_{\bm}(\fq)$ 
in \eqref{hata_string_n} with the exchange 
$N \leftrightarrow n$ are also invariant 
under the outer automorphisms. 
Some string functions are summarized in Appendix \ref{app:string_fn}.

Let us now recall the branching functions of the coset \eqref{coset_A} 
that we refer as the $\mathcal{W}^{\, para}_{N, n}$ characters when 
$p$ is taken to be infinity (or a generic value) and 
the $\mathcal{W}^{\, para}_{N, n}$ $(p,p+n)$-minimal model characters when 
$p$ is an integer with $p \ge N$.
Up to a normalization factor, 
the $\mathcal{W}^{\, para}_{N, n}$ characters 
are given by the $\widehat{\mathfrak{sl}}(N)$ string functions \eqref{gen_string_fn_Nn} of level-$n$ 
(see \cite{Kakushadze:1993jf} for $N=2$), and 
the $\mathcal{W}^{\, para}_{N, n}$ $(p,p+n)$-minimal model characters, 
labelled by 
$\bel=[\ell_0,\ell_1,\ldots,\ell_{N-1}]\in P^{+}_{N,n}$, 
$\br=[r_0,r_1,\ldots,r_{N-1}]\in P^{++}_{N,p}$ and 
$\bs=[s_0,s_1,\ldots,s_{N-1}]\in P^{++}_{N,p+n}$ with 
the non-zero condition
\begin{align}
\sum_{I=1}^{N-1} \left(\ell_I +r_I-s_I\right)\Lambda_I \in 
\overline{Q}_{N},
\quad i.e. \quad
\sum_{I=1}^{N-1} I \, \ell_I \equiv \sum_{I=1}^{N-1} I \, (s_I-r_I)\ \
(\mathrm{mod}\ N),
\label{minimal_condition}
\end{align}
are given by \cite{Christe:1988vc, Bouwknegt:1990fb},
\begin{align}
C_{\,\bel}^{\, \br,\bs}(\fq)=
\mathop{\sum_{\bm\, \in\, P^{+}_{N,n}}}
\limits_{\sum_{I=1}^{N-1} I\, (m_I-\ell_I) \equiv 0 \, (\mathrm{mod}\, N)}
\hat{c}^{\, \bel}_{\,\bm}(\fq) \,
\sum_{w \in \overline{W}}
\sum_{\bk \in K_{w}^{\, \br,\bs}(\bm)} (-1)^{|w|}\, 
\fq^{\, {B}_{p \bk + \br, w(\bs)}-{B}_{\br,\bs}}\,.
\label{br_ch_gen_N}
\end{align}
Here
\begin{align}
K_{w}^{\, \br,\bs}(\bm) = 
\bigcup_{w^{\prime}\in \overline{W}}
\left\{\, \bk \in \overline{Q}_{N} \, \Big|\, 
p \bk + \overline{\br} - w(\overline{\bs}) +
w^{\prime}(\overline{\bm}) \equiv 0\ (\mathrm{mod}\ n\, \overline{Q}_{N})\, \right\}
\label{mm_ch_K}
\end{align}
with
$\overline{\bm}=\sum_{I=1}^{N-1}m_I \Lambda_I$, 
$\overline{\br}=\sum_{I=1}^{N-1}r_I \Lambda_I$, 
$\overline{\bs}=\sum_{I=1}^{N-1}s_I \Lambda_I$, and 
$\overline{W}$ is the finite part of 
the affine Weyl group of $\widehat{\mathfrak{sl}}(N)$,%
\footnote{\,
The Weyl group $\overline{W}$ is generated by the simple Weyl reflections 
$\sfs_I$, $1\le I <N$, acting on 
$\overline{\Lambda}=\sum_{I=1}^{N-1}d_I \Lambda_I$ as 
$\sfs_I(\overline{\Lambda})=\overline{\Lambda}-\inner{\alpha_I}{\overline{\Lambda}}\alpha_I$, \textit{i.e.}
$\sfs_I:$ $d_J \mapsto d_J - \overline{A}_{IJ}d_I$, where 
the simple Weyl reflections have the relations 
$\sfs_I^2=1$ for $1\le I < N$, 
$(\sfs_I \sfs_{I+1})^3=1$ for $1\le I < N-1$, 
$\sfs_I \sfs_J=\sfs_J \sfs_I$ for $|I-J|\ge 2$, and 
$\overline{A}$ is the Cartan matrix of $\mathfrak{sl}(N)$.
}
$|w|$ is the length of $w$, and 
\begin{align}
{B}_{\br, \bs}=
\frac{\left|(p+n)\, \br - p \bs \right|^2}
{2np\,(p+n)}\,.
\end{align}
Note that the formula in \cite{Christe:1988vc, Bouwknegt:1990fb} 
corresponding to \eqref{br_ch_gen_N} has the summation 
over $\bm \in P_{N,n}/n \overline{Q}_{N}$ instead of $\bm \in P^{+}_{N,n}$ 
and the set corresponding to \eqref{mm_ch_K} does have the union 
over $w^{\prime}\in \overline{W}$. 
Here to rewrite it we used the invariance of the string functions 
in footnote \ref{fn:maximal_wt}. We also used the fact that the simple affine 
Weyl reflection $\sfs_0$ on $\Lambda=\sum_{I=0}^{N-1}d_I \Lambda_I \in P_{N,n}$ 
given by $\sfs_0:$ $d_I \mapsto d_I - A_{0I}d_0 \equiv 
d_I - \sum_{J,K=1}^{N-1} A_{IJ}d_K$ $(\mathrm{mod}\ n \overline{Q}_{N})$ 
is also written as 
$\sfs_1\sfs_2 \cdots \sfs_{N-1}\sfs_{N-2}\cdots \sfs_2\sfs_1\in \overline{W}$, 
\textit{i.e.} $\sfs_0 \equiv \sfs_1\sfs_2 \cdots \sfs_{N-1}\sfs_{N-2}\cdots \sfs_2\sfs_1$ on $\Lambda$ modulo $n \overline{Q}_{N}$.

\begin{remark}
Up to a normalization factor, 
the branching function $C_{\,\bel}^{\, \br, \bs}(\fq)$ in \eqref{br_ch_gen_N} 
is defined by
\begin{align}
\chi_{\bel}^{\widehat{\mathfrak{sl}}(N)_n}(\fq, \hat{\bft})\,
\chi_{\br-\boldsymbol{1}}^{\widehat{\mathfrak{sl}}(N)_{p-N}}(\fq, \hat{\bft}) 
\sim
\sum_{\bs \in P^{++}_{N,p+n}}
C_{\,\bel}^{\, \br, \bs}(\fq)\,
\chi_{\bs-\boldsymbol{1}}^{\widehat{\mathfrak{sl}}(N)_{n+p-N}}(\fq, \hat{\bft}),
\label{br_ch_gen_N_def}
\end{align}
where $\boldsymbol{1}=\rho$.
\end{remark}

\begin{exam}\label{ex:n1_w_ch}
When $n=1$, $\mathcal{W}^{\, para}_{N, 1}=\mathcal{W}_{N}$ 
\cite{Zamolodchikov:1985wn, Fateev:1987vh, Fateev:1987zh}. 
The string functions \eqref{gen_string_fn_Nn} for $n=1$ (\textit{i.e.} $\mathcal{W}_{N}$ characters) 
do not depend on the dominant highest-weight 
$\bel \in P^{+}_{N,1}$ and are given by
\begin{align}
\hat{c}(\fq)=\frac{1}{(\fq;\fq)_{\infty}^{N-1}}\,.
\end{align}
Similarly, the $\mathcal{W}_{N}$ $(p,p+1)$-minimal model characters 
\eqref{br_ch_gen_N} for $n=1$ do not depend on the dominant highest-weight 
$\bel \in P^{+}_{N,1}$ and are given by \cite{Mizoguchi:1988pf, Frenkel:1992ju},
\begin{align}
C^{\, \br,\bs}(\fq)=
\frac{1}{(\fq;\fq)_{\infty}^{N-1}}
\sum_{w \in \overline{W}}
\sum_{\bk \in \overline{Q}_{N}} (-1)^{|w|}\, 
\fq^{\, {B}_{p \bk + \br, w(\bs)}-{B}_{\br,\bs}}\,.
\end{align}
\end{exam}

\begin{exam}
When $N=2$, the $\mathcal{W}^{\, para}_{2, n}$ $(p,p+n)$-minimal model characters, labelled by $\bel=[n-\ell,\ell]\in P^{+}_{2,n}$, 
$\br=[p-r,r]\in P^{++}_{2,p}$ and $\bs=[p+n-s,s]\in P^{++}_{2,p+n}$ with 
$\ell+r-s \in 2{\IZ}$, are computed by \cite{Kastor:1987dk, Bagger:1987kv, Ravanini:1987pk},
\begin{align}
C_{\,\bel}^{\, \br, \bs}(\fq)=
\fq^{\, -{B}_{r,s}}
\mathop{\sum_{m=0}^{n}}\limits_{m \equiv \ell\, (\mathrm{mod}\, 2)}
\hat{c}^{\, [n-\ell,\ell]}_{\,[n-m,m]}(\fq)
\left(\mathop{\sum_{k \in {\IZ}}}\limits_{ pk-\frac{r-s}{2} 
\equiv \pm \frac{m}{2} \, (\mathrm{mod}\, n)}
\fq^{\, {B}_{2pk+r,s}} -
\mathop{\sum_{k \in {\IZ}}}\limits_{ pk-\frac{r+s}{2} 
\equiv \pm \frac{m}{2} \, (\mathrm{mod}\, n)}
\fq^{\, {B}_{2pk+r,-s}}\right),
\label{br_ch_gen_2}
\end{align}
where 
$
{B}_{r,s}=
((p+n)r - ps)^2/
(4np(p+n))
$ 
and the string functions $\hat{c}^{\, [n-\ell,\ell]}_{\,[n-m,m]}(\fq)$ are 
given in \eqref{string_fn_N2}.
\end{exam}

\subsection{Dual dominant integral weights}\label{subsec:dual_dw}

For proposing our conjecture, let us define a dominant integral weight
\begin{align}
\bN_{\bel}^{(f)}=
[N_0,N_1,\ldots,N_{n-1}] \in P^{+}_{n,N}
\label{dw_dual}
\end{align}
of $\widehat{\mathfrak{sl}}(n)_N$ which is dual to 
or associated with the dominant integral weight 
$\bel=\left[\ell_0,\ell_1,\ldots,\ell_{N-1}\right]\in P^{+}_{N,n}$ 
of $\widehat{\mathfrak{sl}}(N)_n$. 
Here a non-negative integer 
$f < \mathrm{max}\{N, n\}$, which classifies the dominant weights in 
$P^{+}_{N,n}/\overline{Q}_N$ and $P^{+}_{n,N}/\overline{Q}_n$, 
respectively, as the ${\IZ}_N$ orbits and the ${\IZ}_n$ orbits, 
is introduced by
\footnote{\,
The numbers of the dominant integral weights 
$|P^+_{N,n}|=\frac{(n+N-1)!}{(N-1)!\, n!}$ in $\widehat{\mathfrak{sl}}(N)_n$ and 
$|P^+_{n,N}|=\frac{(n+N-1)!}{(n-1)!\, N!}$ in $\widehat{\mathfrak{sl}}(n)_N$ 
are related by 
$|P^+_{N,n}|/N=|P^+_{n,N}|/n$.
}
\begin{align}
\sum_{I=1}^{N-1} I\, \ell_I \equiv f \ \ (\mathrm{mod}\ N),
\qquad
\sum_{i=1}^{n-1} i\, N_i \equiv f \ \ (\mathrm{mod}\ n).
\label{constraint_hw}
\end{align}
Let $\partit(\bel)=(\lambda_1,\lambda_2,\ldots)$ be the partition 
for $\bel$ in \eqref{partition_L}, and then 
the first relation in \eqref{constraint_hw} is written as 
$\sum_{I=1}^{N-1} \lambda_I \equiv f$ $(\mathrm{mod}\ N)$. 
We define the Dynkin labels $N_{i}$ in \eqref{dw_dual} 
as the multiplicity of $i=\sigma_I^*$ in 
$\{\sigma_1^*,\ldots,\sigma_N^* \}$, where 
$\sigma_I^* \in \{0,1,\ldots,n-1\}$, $1 \le I \le N$, correspond to 
the ${\IZ}_n$ charges (on the gauge side) defined by
\begin{align}
\sigma_I^* \equiv \lambda_I + \sigma_N^* \ \ (\mathrm{mod}\ n),
\ \ 1 \le I < N,
\quad
\sigma_N^* \equiv -\frac{1}{N} \ll 
\sum_{I=1}^{N-1} \lambda_I - f \rr
\ \ (\mathrm{mod}\ n).
\label{sigma_partition}
\end{align}
Here the shifted transposed partition 
$\widetilde{\partit}(\bel)^T=(\lambda_1^T-\lambda_n^T, \lambda_2^T-\lambda_n^T, \ldots)$ by $\lambda_n^T$ naturally defines a 
\textit{\lq transposed\rq} (dual) dominant integral weight 
$\bel^{\,T}=\left[\ell_0^{\,T},\ell_1^{\,T},\ldots,\ell_{n-1}^{\,T}\right]\in P^{+}_{n,N}$ by inverting \eqref{partition_L}. 
Then the first relations in \eqref{sigma_partition} imply that  
the dual dominant integral weight $\bN_{\bel}^{(f)}=[N_0,N_1,\ldots,N_{n-1}]$ 
is defined by
\begin{align}
N_i=\ell_{i-\sigma_N^*}^{\,T},
\ \ \ \ 0 \le i < n,
\end{align}
where we set $\ell_{i+n}^{\,T}=\ell_{i}^{\,T}$. 
Note that, by \eqref{sigma_partition} we see that 
the ${\IZ}_n$ charges $\sigma_I^*$ have the relations
\begin{align}
\sigma_I^* - \sigma_{I+1}^* = \ell_I - n\, \delta_{I,N-g},
\ \ 1 \le I \le N,
\qquad
\sum_{I=1}^{N} \sigma_I^* \equiv f \ \ (\mathrm{mod}\ n),
\label{orbit_charge_suN}
\end{align}
and are ordered as 
$\sigma_{1-g}^* \ge \sigma_{2-g}^* \ge \ldots \ge \sigma_{N-g}^*$, 
where $\sigma_{I+N}^*=\sigma_I^*$, $\ell_{I+N}=\ell_{I}$, 
and $g \in \{0,1,\ldots,N-1\}$ is
\footnote{\,
In terms of the ${\IZ}_n$ charges $\sigma_I=\sigma_{I-g}^*$ with 
the ordering $\sigma_1 \ge \sigma_2 \ge \ldots \ge \sigma_N$, 
the first relations in \eqref{orbit_charge_suN} are written as 
$\sigma_I-\sigma_{I+1}=\ell_{I-g}- n\, \delta_{I,0}$, $0 \le I < N$. 
\label{fn:ordering_charge}}
\begin{align}
g \equiv \frac{1}{n} \ll \sum_{I=1}^{N} \sigma_I^* - f \rr
\equiv \frac{1}{n} \ll \sum_{i=1}^{n-1} i\, N_i - f \rr 
\ \ (\mathrm{mod}\ N).
\label{shift_g}
\end{align}
Here the second relation in \eqref{orbit_charge_suN} gives the second relation 
in \eqref{constraint_hw} by $\sum_{I=1}^N \sigma_I^*=\sum_{i=1}^{n-1} i\, N_i$. 
Some examples of $\bN_{\bel}^{(f)}$ are provided in Appendix \ref{app:hw_ex}.

\begin{figure}[t]
\centering
\includegraphics[width=80mm]{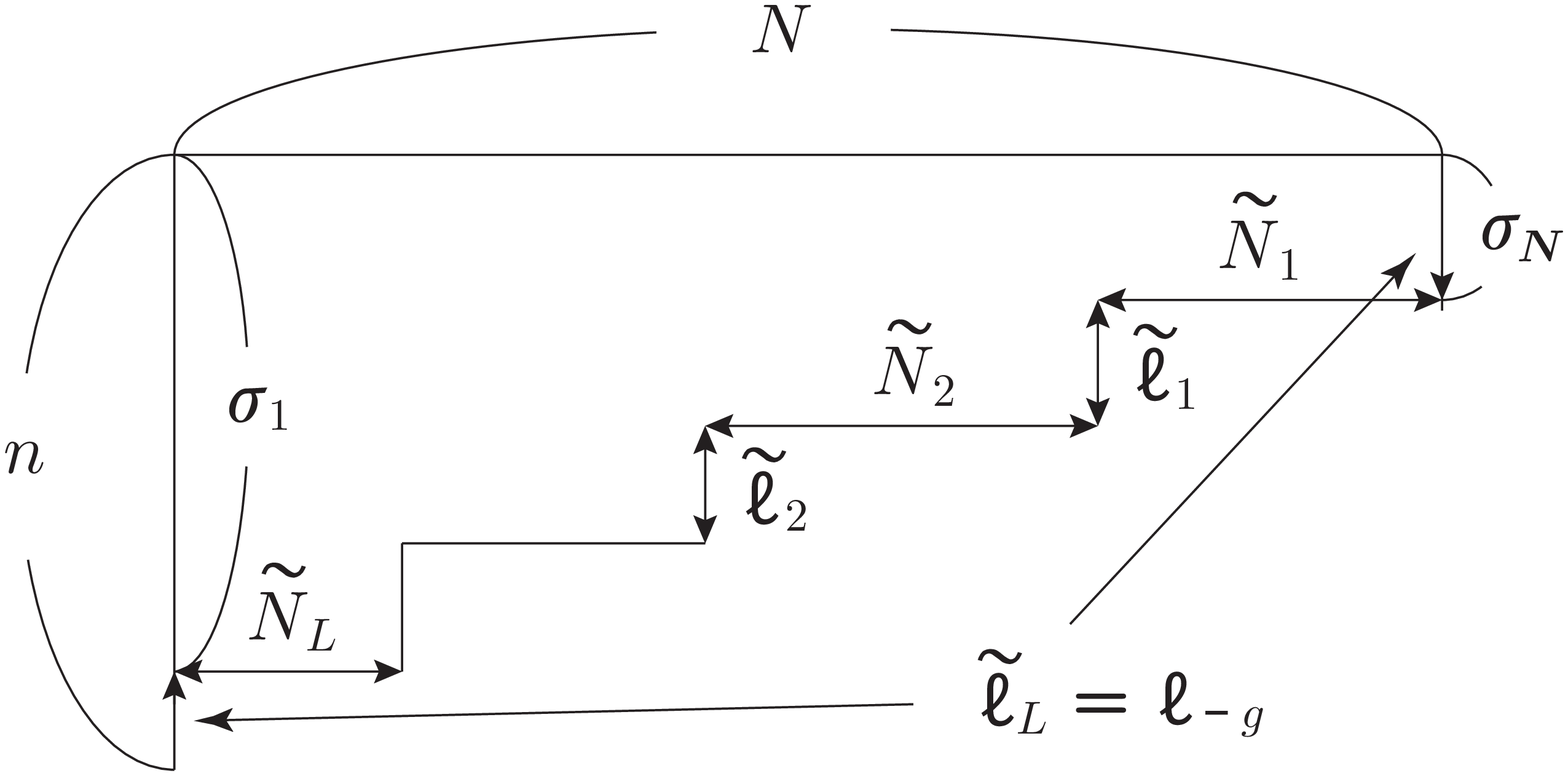}
\caption{The finite sequence \eqref{sequence_rem}, where we follow 
the notation in footnote \ref{fn:ordering_charge}.}
\label{fig:partition_Nn}
\end{figure}

\begin{remark}
Consider the dual dominant integral weight $\bN_{\bel}^{(f)}$. 
Then, the normalization factors of string functions in \eqref{hata_string_n} 
for $\bfc=\bN_{\bel}^{(f)}$ and in \eqref{gen_string_fn_Nn} 
for $\bm=\bel$ are related by
\begin{align}
w_{\bN_{\bel}^{(f)}} - h_{\bN_{\bel}^{(f)}} =
h_{\bel} - \frac{1}{2n}|\bel|^2.
\label{dual_wt_rel}
\end{align}
\begin{proof}
The left and right hand sides are, respectively, obtained as
\begin{align}
w_{\bN_{\bel}^{(f)}} - h_{\bN_{\bel}^{(f)}}=
\frac{1}{2n(n+N)}\sum_{0 \le i < j < n}
(j-i)\, (n-j+i)\, N_i \, N_j\, ,
\label{hw_rem2}
\end{align}
and
\begin{align}
h_{\bel} - \frac{1}{2n}|\bel|^2=
\frac{1}{2n(n+N)}\sum_{0 \le I < J < N}
(J-I)\, (N-J+I)\, \ell_I \, \ell_J\, .
\label{hw_rem1}
\end{align}
We now take all the non-zero components 
$(\widetilde{N}_1, \widetilde{N}_2, \ldots,\widetilde{N}_L)=
(N_{i_1}, N_{i_2}, \ldots, N_{i_L})$ with $i_k < i_{k+1}$ 
from $\bN_{\bel}^{(f)}$, and 
$(\widetilde{\ell}_1, \widetilde{\ell}_2, \ldots,\widetilde{\ell}_L)=
(\ell_{I_1}^{\,\prime}, \ell_{I_2}^{\,\prime}, \ldots, \ell_{I_L}^{\,\prime})$ 
with $I_k > I_{k+1}$ 
from $\bel$, where $\ell_I^{\,\prime}=\ell_{I-g}$, $0 \le I < N$, 
in footnote \ref{fn:ordering_charge}. 
Then consider the finite sequence
\begin{align}
\widetilde{N}_1\, ,\ \widetilde{\ell}_1\, ,\ \widetilde{N}_2\, ,\ 
\widetilde{\ell}_2\, ,\ \ldots,\
\widetilde{N}_L\, ,\ \widetilde{\ell}_L\, ,
\label{sequence_rem}
\end{align}
which is described as in Figure \ref{fig:partition_Nn}. 
For $\widetilde{N}_a=N_{i_a}$, $\widetilde{N}_b=N_{i_b}$ with $a<b$, 
we see that 
$i_b -i_a=\sum_{a \le A < b} \widetilde{\ell}_A$ and 
$n-i_b +i_a=\sum_{A < a} \widetilde{\ell}_A+\sum_{A \ge b} \widetilde{\ell}_A$. 
This shows that \eqref{hw_rem2} is equal to
\begin{align}
\frac{1}{2n(n+N)}
\ll \sum_{1 \le A < a \le B < b \le L}
+\sum_{1 \le a \le A < b \le B \le L} \rr
\widetilde{\ell}_A \, \widetilde{\ell}_B\,
\widetilde{N}_{a}\, \widetilde{N}_{b} \,.
\label{hw_rem12}
\end{align}
Similarly, \eqref{hw_rem1} is also shown to be equal to \eqref{hw_rem12}, 
and thus \eqref{dual_wt_rel} is proved.
\end{proof}
\end{remark}

When we consider the special case $\epsilon_1+\epsilon_2=0$ ($p\to \infty$), 
the central charge \eqref{cc_p_inf} of the AGT-corresponding CFT 
is reminiscent of a conformal embedding
\begin{align}
\mathcal{H} \oplus 
\widehat{\mathfrak{sl}}(n)_N \oplus \widehat{\mathfrak{sl}}(N)_n 
\subset
\widehat{\mathfrak{gl}}(Nn)_1,
\label{conf_embed}
\end{align}
which preserves the central charge $Nn$ and 
is utilized to explain the level-rank duality between 
$\widehat{\mathfrak{sl}}(n)_N$ and $\widehat{\mathfrak{sl}}(N)_n$, 
where $\widehat{\mathfrak{gl}}(Nn)_1 \cong 
\mathcal{H}^{\oplus N} \oplus \widehat{\mathfrak{sl}}(n)_1^{\oplus N}$ is described by $Nn$ free fermions 
\cite{Frenkel:1982, JimboMiwa:85, Hasegawa:89, Naculich:1990hg, Nakanishi:1990hj}
(see also \cite{Dijkgraaf:2007sw} for an elegant string theory interpretation 
by intersecting D4 and D6-branes). 
Actually, the generating function \eqref{inst_ch} 
of coloured Young diagrams for general $N$ and $n$ is obtained by 
\cite{Fujii:2005dk, Alfimov:2013cqa}
\begin{align}
\widehat{X}_{\bsig;\bdel \bk}(\fq)=
\frac{1}{\left( \fq;\fq \right)_{\infty}^{N-1}}\, 
\sum_{\bdel \bk_1 + \cdots + \bdel \bk_N =\bdel \bk}
\prod_{I=1}^N \widehat{X}_{(\sigma_I);\bdel \bk_I}(\fq)\,,
\label{formula_gen_fn}
\end{align}
where 
\begin{align}
\widehat{X}_{(\sigma);\bdel \bk}(\fq) =
\frac{1}{\left( \fq;\fq \right)_{\infty}^{n-1}}\, 
\fq^{\sum_{i=1}^{n-1} \left(\delta k_{i}^2 + \frac{\delta k_{i}}{n} 
-\delta k_{i-1}\, \delta k_{i} - \delta_{\sigma i}\, \delta k_{i}\right)},
\end{align}
is the generating function for $N=1$ 
which gives the $\widehat{\mathfrak{sl}}(n)_1$ WZW character. 
Let $\bsig_{\bm}^{(f)}=(\sigma_1,\ldots,\sigma_N)$ be 
the ${\IZ}_n$ charges 
with the ordering $\sigma_1 \ge \sigma_2 \ge \ldots \ge \sigma_N$ 
which follow from the dual dominant integral weight $\bN_{\bm}^{(f)}$ in \eqref{dw_dual}. 
Following the algebra $\mathcal{A}(N,n;p)$ for $p\to \infty$, 
we find that the generating function \eqref{formula_gen_fn} is decomposed into 
the $\widehat{\mathfrak{sl}}(N)$ string functions \eqref{gen_string_fn_Nn} 
of level-$n$ and 
the $\widehat{\mathfrak{sl}}(n)$ string functions \eqref{hata_string_n} 
of level-$N$ as
\begin{align}
\begin{split}
\widehat{X}_{\bsig_{\bm}^{(f)};\bdel \bk}(\fq)&=
\mathop{\sum_{\bel\, \in\, P^{+}_{N,n}}}
\limits_{\sum_{I=1}^{N-1} I\, \ell_I \equiv f \, (\mathrm{mod}\, N)}
\hat{c}^{\, \bel}_{\,\bm}(\fq) \times 
\hat{a}^{\bN_{\bel}^{(f)}}_{\bfc(\bdel \bk)}(\fq)
\\
&=
\mathop{\sum_{\bel\, \in\, P^{+}_{N,n}}}
\limits_{\sum_{I=1}^{N-1} I\, \ell_I \equiv f \, (\mathrm{mod}\, N)}
\fq^{\, \frac{1}{2n}\, (|\bel|^2-|\bm|^2)
+\frac{1}{n}\sum_{i=1}^{n-1}\delta k_i}\, 
a^{\, \bel}_{\,\bm}(\fq) \times 
a^{\bN_{\bel}^{(f)}}_{\bfc(\bdel \bk)}(\fq)\,,
\label{eq:conj_string}
\end{split}
\end{align}
where, in the second equality the relation \eqref{dual_wt_rel} was used. 
The same decomposition was shown for the conformal embedding 
\eqref{conf_embed} in \cite{Hasegawa:89} 
(see also \cite[Appendix A]{Dijkgraaf:2007sw}). 
In terms of the $SU(N)$ $\ft$-refined generating functions 
\eqref{inst_ch_gen} of $n$-coloured Young diagrams, the above decomposition 
boils down to the decomposition into 
the $\widehat{\mathfrak{sl}}(N)$ string functions \eqref{gen_string_fn_Nn} 
of level-$n$ ($\mathcal{W}^{\, para}_{N, n}$ characters) and 
the $\widehat{\mathfrak{sl}}(n)_N$ WZW characters \eqref{wzw_ch} as
\begin{align}
\widehat{X}_{\bN_{\bm}^{(f)}}(\fq, \bft)=
\mathop{\sum_{\bel\, \in\, P^{+}_{N,n}}}
\limits_{\sum_{I=1}^{N-1} I\, \ell_I \equiv f \, (\mathrm{mod}\, N)}
\hat{c}^{\, \bel}_{\,\bm}(\fq) \times
\chi_{\bN_{\bel}^{(f)}}^{\widehat{\mathfrak{sl}}(n)_N}(\fq, \hat{\bft})\,,
\label{eq:decomp_ch}
\end{align}
where $\hat{\ft}_{i}=\fq^{- \frac{i\, (n-i)}{2n}}\, \ft_i$.

\subsection{Conjecture}

Based on the symmetry algebra $\mathcal{A}(N,n;p)$ in \eqref{alg_A}, 
we now propose the following conjecture for integers $p \ge N$ that 
generalizes the decomposition formula \eqref{eq:decomp_ch} for $p \to \infty$.

\begin{conj}\label{conj:suN_red_ch}
The $SU(N)$ $\ft$-refined Burge-reduced generating functions \eqref{rs_red_ch_gen} of $n$-coloured Young diagrams 
can be decomposed into 
the $\mathcal{W}^{\, para}_{N, n}$ $(p,p+n)$-minimal model 
characters \eqref{br_ch_gen_N} and 
the $\widehat{\mathfrak{sl}}(n)_N$ WZW characters \eqref{wzw_ch} as
\begin{align}
\widehat{X}_{\bN_{\bm}^{(f)}}^{\, \br,\bs}(\fq, \bft)=
\mathop{\sum_{\bel\, \in\, P^{+}_{N,n}}}
\limits_{\sum_{I=1}^{N-1} I\, \ell_I \equiv f \, (\mathrm{mod}\, N)}
C_{\,\bel^{\,(\omega)}}^{\, \br,\bs}(\fq) \times
\chi_{\bN_{\bel}^{(f)}}^{\widehat{\mathfrak{sl}}(n)_N}(\fq, \hat{\bft})\,,
\label{eq:conj_minimal}
\end{align}
where $\hat{\ft}_{i}=\fq^{- \frac{i\, (n-i)}{2n}}\, \ft_i$. 
The dominant weight $\bel^{\,(\omega)}
=[\ell^{\,(\omega)}_0, \ell^{\,(\omega)}_1, \ldots, \ell^{\,(\omega)}_{N-1}] \in P^{+}_{N,n}$ 
with $\ell^{\,(\omega)}_{I}=\ell_{I-\omega}$, $\ell_{I+N}=\ell_{I}$, 
$0 \le I < N$, is shifted by
\begin{align}
\omega \equiv
\frac{1}{n}\ll \sum_{i=1}^{n-1} i\, N_i - f \rr + \frac{1}{n} \, 
\sum_{I=1}^{N-1} I \ll s_I - r_I - \sigma_I + \sigma_{I+1} \rr
\ \ (\mathrm{mod}\ N),
\label{theta_conj}
\end{align}
where $\bsig_{\bm}^{(f)}=(\sigma_1,\ldots,\sigma_N)$ 
are the ${\IZ}_n$ charges associated with $\bN_{\bm}^{(f)}=[N_0,N_1,\ldots,N_{n-1}] \in P^{+}_{n,N}$, 
and the ordering of $\bsig_{\bm}^{(f)}$ depends on $\br$ and $\bs$ 
by the ${\IZ}_n$ charge conditions \eqref{charge_condition}. 
Here the non-zero condition \eqref{minimal_condition} for 
the characters $C_{\,\bel^{\,(\omega)}}^{\, \br,\bs}(\fq)$ 
is shown to be satisfied as
\begin{align}
\sum_{I=1}^{N-1} I \, \ell^{\,(\omega)}_I \equiv
\sum_{I=1}^{N-1} I \, \ell_I + \omega\, n \equiv
\sum_{I=1}^{N-1} I \, (s_I - r_I) \ \ 
(\mathrm{mod}\ N),
\label{ellhat}
\end{align}
where in the second equality we used 
$\sum_{I=1}^{N-1} I \ell_I \equiv f$ 
and $\sum_{I=1}^{N-1} I (\sigma_I - \sigma_{I+1}) 
\equiv \sum_{i=1}^{n-1}i\, N_i$ $(\mathrm{mod}\ N)$. 
By the expansions \eqref{rs_red_ch_gen} and \eqref{wzw_string}, 
the conjectural formula \eqref{eq:conj_minimal} is equivalent to 
\begin{align}
\widehat{X}_{\bsig_{\bm}^{(f)};\bdel \bk}^{\, \br,\bs}(\fq)=
\mathop{\sum_{\bel\, \in\, P^{+}_{N,n}}}
\limits_{\sum_{I=1}^{N-1} I\, \ell_I \equiv f \, (\mathrm{mod}\, N)}
C_{\,\bel^{\,(\omega)}}^{\, \br,\bs}(\fq) \times 
\hat{a}^{\bN_{\bel}^{(f)}}_{\bfc(\bdel \bk)}(\fq)\,.
\label{eq:conj_minimal_c}
\end{align}
\end{conj}

We make some remarks to support Conjecture \ref{conj:suN_red_ch}.

\begin{remark}
From the invariance \eqref{inv_b_red_g} of 
the $\ft$-refined Burge-reduced generating functions 
under the cyclic permutations $\sigma_I \to \sigma_{I-\theta}$, 
$r_I \to r^{(\theta)}=r_{I-\theta}$ and $s_I \to s^{(\theta)}=s_{I-\theta}$, 
$\theta \in {\IZ}_N$, one finds that 
the conjectural formula \eqref{eq:conj_minimal} gives a relation
\begin{align}
\mathop{\sum_{\bel\, \in\, P^{+}_{N,n}}}
\limits_{\sum_{I=1}^{N-1} I\, \ell_I \equiv f \, (\mathrm{mod}\, N)}
\ll C_{\,\bel^{\,(\omega)}}^{\, \br,\bs}(\fq)
- C_{\,\bel^{\,(\omega+\theta)}}^{\, \br^{(\theta)},\bs^{(\theta)}}(\fq) \rr
\chi_{\bN_{\bel}^{(f)}}^{\widehat{\mathfrak{sl}}(n)_N}(\fq, \hat{\bft})=0.
\label{conj_ch_sym}
\end{align}
Here $\omega$ is defined by \eqref{theta_conj} and 
\begin{align}
\frac{1}{n} \, 
\sum_{I=1}^{N-1} I \ll s^{(\theta)}_I - r^{(\theta)}_I - \sigma_{I-\theta} + \sigma_{I-\theta+1} \rr 
\equiv
\frac{1}{n} \, 
\sum_{I=1}^{N-1} I \ll s_I - r_I - \sigma_I + \sigma_{I+1} \rr
+ \theta
\ \ (\mathrm{mod}\ N),
\end{align}
is used. The relation \eqref{conj_ch_sym} then implies the invariance
\begin{align}
C_{\,\bel}^{\, \br,\bs}(\fq)
= C_{\,\bel^{\,(\theta)}}^{\, \br^{(\theta)},\bs^{(\theta)}}(\fq),
\quad
\theta \in {\IZ}_N,
\end{align}
of the minimal model characters (branching functions).
\end{remark}

\begin{remark}\label{rm:sp_reduce}
In the special case $p=N$, let us show that 
the conjectural formula (\ref{eq:conj_minimal}) yields 
the formula (\ref{wzw_ch_comb_c}). In this special case, by
\begin{align}
C_{\,\bel}^{\, \boldsymbol{1},\bs}(\fq)=
\begin{cases}
\fq^{\, h_{\bel} - \frac{1}{2n}|\bel|^2}
&\textit{if}\ \ 
s_I=\ell_I+1\ \ \textit{for}\ \ 0 \le I <N,
\\
0
&\textit{otherwise},
\end{cases}
\end{align}
which follows from the definition \eqref{br_ch_gen_N_def} 
with taking into account of the normalization factor, 
the conjectural formula \eqref{eq:conj_minimal} is
\begin{align}
\widehat{X}_{\bN_{\bm}^{(f)}}^{\, \boldsymbol{1},\bs}(\fq, \bft)=
\mathop{\sum_{\bel\, \in\, P^{+}_{N,n}}}
\limits_{\sum_{I=1}^{N-1} I\, \ell_I \equiv f \, (\mathrm{mod}\, N)}
\fq^{\, h_{\bel^{\,(\omega)}} - \frac{1}{2n}|\bel^{\,(\omega)}|^2}\,
\chi_{\bN_{\bel}^{(f)}}^{\widehat{\mathfrak{sl}}(n)_N}(\fq, \hat{\bft})\,
\prod_{I=0}^{N-1}\delta_{\ell^{\,(\omega)}_{I},\, s_I-1}\,.
\label{eq:conj_sp_c}
\end{align}
Following footnote \ref{fn:sp_wzw_ch}, by taking 
$\sigma_1 \ge \sigma_2 \ge \ldots \ge \sigma_N$ and 
$\sigma_I-\sigma_{I+1}=s_I-1-n\, \delta_{I,0}$, 
the shift parameter $\omega$ is now given by $\omega=g$ in \eqref{shift_g} 
and then $m_I^{(g)}=s_I-1$ by footnote \ref{fn:ordering_charge}, 
where note that the ${\IZ}_n$ charges $\sigma_I$ 
are associated with $\bN_{\bm}^{(f)}$. 
As a result, \eqref{eq:conj_sp_c} yields
\begin{align}
\widehat{X}_{\bN_{\bm}^{(f)}}^{\, \boldsymbol{1},\bs}(\fq, \bft)=
\fq^{\, h_{\bm^{(g)}} - \frac{1}{2n}|\bm^{(g)}|^2}\,
\chi_{\bN_{\bm}^{(f)}}^{\widehat{\mathfrak{sl}}(n)_N}(\fq, \hat{\bft})\,.
\label{eq:conj_sp}
\end{align}
Therefore, by 
$
h_{\bm^{(g)}} - \frac{1}{2n}|\bm^{(g)}|^2
=h_{\bm} - \frac{1}{2n}|\bm|^2 
$
following from \eqref{hw_rem1}, and by the relation \eqref{dual_wt_rel} 
we obtain the formula (\ref{wzw_ch_comb_c}). 
\end{remark}

\begin{remark}
When $n=1$, the conjectural formula \eqref{eq:conj_minimal} yields 
\begin{align}
\widehat{X}_{[N]}^{\, \br,\bs}(\fq)=
C^{\, \br,\bs}(\fq),
\end{align}
which gives the $\mathcal{W}_{N}$ $(p,p+1)$-minimal model characters in 
Example \ref{ex:n1_w_ch}.%
\footnote{\,
See \cite[Section 3.4]{Foda:2015bsa}, where 
note our normalization of string functions as below \eqref{gen_string_fn_Nn} 
by $\frac{1}{24}\, c(\widehat{\mathfrak{sl}}(N)_1)=\frac{1}{24}\, (N-1)$.}
\end{remark}

\section{Examples of Burge-reduced generating functions}\label{sec:examples}

\textit{\noindent
In this section, we test Conjecture \ref{conj:suN_red_ch} by 
extracting the $\mathcal{W}^{\, para}_{N, n}$ $(p,p+n)$-minimal model 
characters from 
the $SU(N)$ Burge-reduced generating functions of $n$-coloured Young diagrams 
in the cases of $(N,n,p)=(2,2,4)$ and $(3,3,4)$. 
By assuming the formula \eqref{eq:conj_minimal_c} 
with the use of the $\widehat{\mathfrak{sl}}(n)$ string functions 
in Appendix \ref{app:string_fn} we will check that 
the minimal model characters in \eqref{br_ch_gen_N} are obtained.}

\subsection{$(N,n)=(2,2)$ and minimal super-Virasoro characters}

When $(N,n)=(2,2)$, the $\mathcal{W}^{\, para}_{2, 2}$ algebra is 
the super-Virasoro algebra \cite{Goddard:1986ee} and 
studied in the context of the AGT correspondence in 
\cite{Belavin:2011pp, Bonelli:2011jx, Belavin:2011tb, Bonelli:2011kv, Ito:2011mw, Belavin:2012aa}.
Here we consider the $(4,6)$-minimal model ($p=4$) which has 
central charge $c\left(\mathcal{W}^{\, para}_{2, 2}\right)=1$ by \eqref{cc_paraW}. 
The $SU(2)$ Burge-reduced generating functions 
$\widehat{X}_{\bsig;(\delta k)}^{\, \br,\bs}(\fq)$ in \eqref{rs_red_ch} of 
$2$-coloured Young diagrams are labelled by 
$\bsig=(\sigma_1,\sigma_2)$ with $0 \le \sigma_1, \sigma_2 \le 1$, 
$\delta k \in {\IZ}$, and 
$\br=[r_0,r_1]\in P^{++}_{2,p}$, $\bs=[s_0,s_1]\in P^{++}_{2,p+2}$
with $s_1 - r_1 \equiv \sigma_1-\sigma_2$ $(\mathrm{mod}\ 2)$, 
where $\bsig$ and $\delta k$ define $\bfc=[\fc_0,\fc_1]=[N_0+2\delta k, N_1-2\delta k]\in P_{2,2}$ in \eqref{inst_chern}.

The Burge-reduced generating functions for $\bN=[2,0]$ and 
$\bfc=[2,0]$, $[0,2]$ are obtained as
\begin{align}
\begin{split}
\widehat{X}_{(0,0);(0)}^{\,[3,1],[5,1]}(\fq)&=
1+\fq+5 \fq^2+10 \fq^3+25 \fq^4+48 \fq^{5}+101 \fq^{6}+185 \fq^{7}+350 \fq^{8}+615 \fq^{9}+\cdots \, ,
\\
\widehat{X}_{(0,0);(-1)}^{\,[3,1],[5,1]}(\fq)&=
\fq^{\frac12}+3 \fq^{\frac32}+7 \fq^{\frac52}+16 \fq^{\frac72}+35 \fq^{\frac92}+70 \fq^{\frac{11}{2}}+137 \fq^{\frac{13}{2}}+256 \fq^{\frac{15}{2}}+465 \fq^{\frac{17}{2}}+\cdots \, ,
\\
\widehat{X}_{(0,0);(0)}^{\,[2,2],[4,2]}(\fq)&=
1+3 \fq+10 \fq^2+25 \fq^3+57 \fq^4+121 \fq^{5}+243 \fq^{6}+465 \fq^{7}+862 \fq^{8}+\cdots \, ,
\\
\widehat{X}_{(0,0);(-1)}^{\,[2,2],[4,2]}(\fq)&=
2 \fq^{\frac12}+6 \fq^{\frac32}+16 \fq^{\frac52}+38 \fq^{\frac72}+84 \fq^{\frac92}+172 \fq^{\frac{11}{2}}+338 \fq^{\frac{13}{2}}+636 \fq^{\frac{15}{2}}+\cdots \, ,
\end{split}
\end{align}
and using the $\widehat{\mathfrak{sl}}(2)$ string functions \eqref{string_fn_N2} 
of level-$2$ with $\hat{a}^{\,[N_0,N_1]}_{\,[\fc_0,\fc_1]}(\fq)=
\fq^{\,\frac{1}{8}\fc_1(\fc_1-2)}\,\hat{c}^{\, [N_0,N_1]}_{\,[\fc_0,\fc_1]}(\fq)$, 
from the formula \eqref{eq:conj_minimal_c} we obtain
\begin{align}
\begin{split}
C_{\,[2,0]}^{\, [3,1],[5,1]}(\fq)&=
1+\fq^2+\fq^3+3 \fq^4+3 \fq^{5}+7 \fq^{6}+8 \fq^{7}+14 \fq^{8}+17 \fq^{9}+27 \fq^{10}+\cdots\,,
\\
C_{\,[0,2]}^{\, [3,1],[5,1]}(\fq)&=
\fq^{\frac32}+\fq^{\frac52}+2 \fq^{\frac72}+3 \fq^{\frac92}+5 \fq^{\frac{11}{2}}+7 \fq^{\frac{13}{2}}+11 \fq^{\frac{15}{2}}+15 \fq^{\frac{17}{2}}+22 \fq^{\frac{19}{2}}+\cdots\,,
\\
C_{\,[2,0]}^{\, [2,2],[4,2]}(\fq)&=
1+\fq+2 \fq^2+4 \fq^3+6 \fq^4+10 \fq^{5}+15 \fq^{6}+22 \fq^{7}+32 \fq^{8}+46 \fq^{9}+\cdots\,,
\\
C_{\,[0,2]}^{\, [2,2],[4,2]}(\fq)&=
\fq^{\frac12}+2 \fq^{\frac32}+3 \fq^{\frac52}+5 \fq^{\frac72}+8 \fq^{\frac92}+12 \fq^{\frac{11}{2}}+18 \fq^{\frac{13}{2}}+27 \fq^{\frac{15}{2}}+38 \fq^{\frac{17}{2}}+\cdots\,.
\end{split}
\end{align}
We see that they agree with the $\mathcal{W}^{\, para}_{2, 2}$ $(4,6)$-minimal model characters in \eqref{br_ch_gen_2}. 
Similarly, from a Burge-reduced generating function for $\bN=[1,1]$ and 
$\bfc=[1,1]$,
\begin{align}
\widehat{X}_{(1,0);(0)}^{\,[3,1],[4,2]}(\fq)&=
1+3 \fq+8 \fq^2+20 \fq^3+44 \fq^4+92 \fq^{5}+183 \fq^{6}+348 \fq^{7}+640 \fq^{8}+1144 \fq^{9}+\cdots \, ,
\end{align}
we obtain
\begin{align}
C_{\,[1,1]}^{\, [3,1],[4,2]}(\fq)&=
\fq^{\frac{1}{16}}\,(1+\fq+2 \fq^{2}+4 \fq^{3}+6 \fq^{4}+10 \fq^{5}+15 \fq^{6}+22 \fq^{7}+32 \fq^{8}+46 \fq^{9}+\cdots)\,.
\end{align}

\subsection{$(N,n)=(3,3)$ and minimal super-$\mathcal{W}_3$ characters}

When $(N,n)=(3,3)$, the $\mathcal{W}^{\, para}_{3, 3}$ algebra is 
supposed to be the super-$\mathcal{W}_3$ algebra, and here we consider 
the $(4,7)$-minimal model $(p=4)$ with the central charge 
$c\left(\mathcal{W}^{\, para}_{3, 3}\right)=10/7$ by \eqref{cc_paraW} 
which ensures the associativity of 
the $\mathcal{W}^{\, para}_{3, 3}$ algebra and has a unitary representation 
\cite{Inami:1988xy, Bilal:1990dw, Hornfeck:1990zw, Ahn:1990nr}.%
\footnote{\,
See \cite{Schoutens:1990xg, Hornfeck:1992he} for the generalization to 
the minimal super-$\mathcal{W}_N$ algebra corresponding to $(N,n,p)=(N,N,N+1)$.}
The $SU(3)$ Burge-reduced generating functions 
$\widehat{X}_{\bsig;\bdel \bk}^{\, \br,\bs}(\fq)$ of 
$3$-coloured Young diagrams are labelled by 
$\bsig=(\sigma_1,\sigma_2,\sigma_3)$ with 
$0 \le \sigma_1, \sigma_2,\sigma_3 \le 2$, 
$\bdel \bk=(\delta k_1,\delta k_2) \in {\IZ}^2$, and 
$\br=[r_0,r_1,r_2]\in P^{++}_{3,4}$, $\bs=[s_0,s_1,s_3]\in P^{++}_{3,7}$ 
with $s_1 - r_1 \equiv \sigma_1-\sigma_2$,  
$s_2 - r_2 \equiv \sigma_2-\sigma_3$ $(\mathrm{mod}\ 3)$, 
where $\bsig$ and $\bdel \bk$ define 
$\bfc=[\fc_0,\fc_1,\fc_2]\in P_{3,3}$ in \eqref{inst_chern}.

The Burge-reduced generating functions for $\bN=[3,0,0]$ and 
$\bfc=[3,0,0]$, $[1,1,1]$, $[0,3,0]$, $[0,0,3]$ are obtained as
\begin{align}
\begin{split}
\widehat{X}_{(0,0,0);(0,0)}^{\,[2,1,1],[5,1,1]}(\fq)&=
1+2 \fq+11 \fq^2+42 \fq^3+144 \fq^{4}+448 \fq^{5}+1303 \fq^{6}+3510 \fq^{7}+\cdots \, ,
\\
\widehat{X}_{(0,0,0);(-1,-1)}^{\,[2,1,1],[5,1,1]}(\fq)&=
\fq^{\frac13}+5 \fq^{\frac43}+24 \fq^{\frac73}+89 \fq^{\frac{10}{3}}+299 \fq^{\frac{13}{3}}+896 \fq^{\frac{16}{3}}+2503 \fq^{\frac{19}{3}}+\cdots \, ,
\\
\widehat{X}_{(0,0,0);(-2,-1)}^{\,[2,1,1],[5,1,1]}(\fq)&=
\fq+8 \fq^2+35 \fq^3+132 \fq^{4}+426 \fq^{5}+1261 \fq^{6}+3443 \fq^{7}+\cdots \, ,
\\
\widehat{X}_{(0,0,0);(-1,-2)}^{\,[2,1,1],[5,1,1]}(\fq)&=
\fq+8 \fq^2+35 \fq^3+132 \fq^{4}+426 \fq^{5}+1261 \fq^{6}+3443 \fq^{7}+\cdots 
\, ,
\end{split}
\end{align}
and using the $\widehat{\mathfrak{sl}}(3)$ string functions \eqref{string_fn_N3_l3} of level-3 with $\hat{a}^{\,\bN}_{\,\bfc}(\fq)=
\fq^{\,\frac{1}{9}(\fc_1^2+\fc_2^2+\fc_1\fc_2)-\frac13(\fc_1+\fc_2)}\,\hat{c}^{\, \bN}_{\,\bfc}(\fq)$, 
from the formula \eqref{eq:conj_minimal_c} we find 
the $\mathcal{W}^{\, para}_{3, 3}$ $(4,7)$-minimal model characters
\begin{align}
\begin{split}
C_{\,[3,0,0]}^{\, [2,1,1],[5,1,1]}(\fq)&=
1+\fq^2+2 \fq^3+3 \fq^{4}+4 \fq^{5}+8 \fq^{6}+10 \fq^{7}+\cdots\,,
\\
C_{\,[1,1,1]}^{\, [2,1,1],[5,1,1]}(\fq)&=
\fq^{\frac32}+2 \fq^{\frac52}+3 \fq^{\frac72}+6 \fq^{\frac92}+10 \fq^{\frac{11}{2}}+16 \fq^{\frac{13}{2}}+\cdots\,,
\\
C_{\,[0,0,3]}^{\, [2,1,1],[5,1,1]}(\fq)&=
\fq^{4}+\fq^{5}+3 \fq^{6}+5 \fq^{7}+\cdots\,,
\\
C_{\,[0,3,0]}^{\, [2,1,1],[5,1,1]}(\fq)&=
\fq^{4}+\fq^{5}+3 \fq^{6}+5 \fq^{7}+\cdots\,.
\end{split}
\end{align}
Similarly, 
the Burge-reduced generating functions for $\bN=[1,1,1]$ and 
$\bfc=[3,0,0]$, $[1,1,1]$, $[0,3,0]$, $[0,0,3]$,%
\footnote{\,
Note the ordering $\sigma_1 \le \sigma_2 \le \sigma_3$ for 
$(\sigma_1, \sigma_2, \sigma_3)=(0,1,2)$.}
\begin{align}
\begin{split}
\widehat{X}_{(0,1,2);(1,1)}^{\,[2,1,1],[1,3,3]}(\fq)&=
3 \fq^{\frac23}+18 \fq^{\frac53}+84 \fq^{\frac83}+312 \fq^{\frac{11}{3}}+1028 \fq^{\frac{14}{3}}+3052 \fq^{\frac{17}{3}}+8425 \fq^{\frac{20}{3}}+\cdots\,,
\\
\widehat{X}_{(0,1,2);(0,0)}^{\,[2,1,1],[1,3,3]}(\fq)&=
1+10 \fq+50 \fq^2+203 \fq^3+693 \fq^{4}+2136 \fq^{5}+6031 \fq^{6}+15967 \fq^{7}+\cdots\,,
\\
\widehat{X}_{(0,1,2);(-1,0)}^{\,[2,1,1],[1,3,3]}(\fq)&=
2 \fq^{\frac23}+16 \fq^{\frac53}+79 \fq^{\frac83}+302 \fq^{\frac{11}{3}}+1009 \fq^{\frac{14}{3}}+3018 \fq^{\frac{17}{3}}+8364 \fq^{\frac{20}{3}}+\cdots\,,
\\
\widehat{X}_{(0,1,2);(0,-1)}^{\,[2,1,1],[1,3,3]}(\fq)&=
3 \fq^{\frac23}+18 \fq^{\frac53}+84 \fq^{\frac83}+312 \fq^{\frac{11}{3}}+1028 \fq^{\frac{14}{3}}+3052 \fq^{\frac{17}{3}}+8425 \fq^{\frac{20}{3}}+\cdots\,,
\end{split}
\end{align}
give
\begin{align}
\begin{split}
C_{\,[3,0,0]}^{\, [2,1,1],[1,3,3]}(\fq)&=
\fq^{\frac83}+2 \fq^{\frac{11}{3}}+5 \fq^{\frac{14}{3}}+8 \fq^{\frac{17}{3}}+15 \fq^{\frac{20}{3}}+\cdots\,,
\\
C_{\,[1,1,1]}^{\, [2,1,1],[1,3,3]}(\fq)&=
\fq^{\frac16}+2 \fq^{\frac76}+4 \fq^{\frac{13}{6}}+8 \fq^{\frac{19}{6}}+15 \fq^{\frac{25}{6}}+26 \fq^{\frac{31}{6}}+43 \fq^{\frac{37}{6}}+\cdots\,,
\\
C_{\,[0,0,3]}^{\, [2,1,1],[1,3,3]}(\fq)&=
\fq^{\frac23}+\fq^{\frac53}+3 \fq^{\frac83}+4 \fq^{\frac{11}{3}}+8 \fq^{\frac{14}{3}}+12 \fq^{\frac{17}{3}}+21 \fq^{\frac{20}{3}}+\cdots\,,
\\
C_{\,[0,3,0]}^{\, [2,1,1],[1,3,3]}(\fq)&=
\fq^{\frac23}+\fq^{\frac53}+3 \fq^{\frac83}+4 \fq^{\frac{11}{3}}+8 \fq^{\frac{14}{3}}+12 \fq^{\frac{17}{3}}+21 \fq^{\frac{20}{3}}+\cdots\,.
\end{split}
\end{align}
The Burge-reduced generating functions for $\bN=[2,1,0]$ and 
$\bfc=[2,1,0]$, $[0,2,1]$, $[1,0,2]$,
\begin{align}
\begin{split}
\widehat{X}_{(1,0,0);(0,0)}^{\,[2,1,1],[4,2,1]}(\fq)&=
1+5 \fq+26 \fq^2+104 \fq^3+367 \fq^{4}+1151 \fq^{5}+3329 \fq^{6}+8969 \fq^{7}+\cdots\,,
\\
\widehat{X}_{(1,0,0);(-1,-1)}^{\,[2,1,1],[4,2,1]}(\fq)&=
\fq^{\frac13}+8 \fq^{\frac43}+39 \fq^{\frac73}+156 \fq^{\frac{10}{3}}+532 \fq^{\frac{13}{3}}+1638 \fq^{\frac{16}{3}}+4631 \fq^{\frac{19}{3}}+\cdots\,,
\\
\widehat{X}_{(1,0,0);(0,-1)}^{\,[2,1,1],[4,2,1]}(\fq)&=
2 \fq^{\frac23}+13 \fq^{\frac53}+62 \fq^{\frac83}+234 \fq^{\frac{11}{3}}+777 \fq^{\frac{14}{3}}+2322 \fq^{\frac{17}{3}}+6435 \fq^{\frac{20}{3}}+\cdots\,,
\end{split}
\end{align}
give
\begin{align}
\begin{split}
C_{\,[2,1,0]}^{\, [2,1,1],[4,2,1]}(\fq)&=
\fq^{\frac19}\,
(1+\fq+3 \fq^2+5 \fq^3+9 \fq^{4}+14 \fq^{5}+24 \fq^{6}+37 \fq^{7}+\cdots)\,,
\\
C_{\,[1,0,2]}^{\, [2,1,1],[4,2,1]}(\fq)&=
\fq^{\frac19}\,
(\fq^{\frac43}+2 \fq^{\frac73}+4 \fq^{\frac{10}{3}}+7 \fq^{\frac{13}{3}}+13 \fq^{\frac{16}{3}}+21 \fq^{\frac{19}{3}}+\cdots)\,,
\\
C_{\,[0,2,1]}^{\, [2,1,1],[4,2,1]}(\fq)&=
\fq^{\frac19}\,
(\fq^{\frac53}+2 \fq^{\frac83}+4 \fq^{\frac{11}{3}}+8 \fq^{\frac{14}{3}}+14 \fq^{\frac{17}{3}}+24 \fq^{\frac{20}{3}}+\cdots)\,.
\end{split}
\end{align}

\section{Summary and outlook}\label{sec:summary}

Following the AGT correspondence for $U(N)$ gauge theory 
on ${\IC}^2/{\IZ}_n$, we conjectured 
the decomposition formula \eqref{eq:conj_minimal} 
of the Burge-reduced generating functions of $N$-tuples of 
$n$-coloured Young diagrams with 
the Burge conditions and the ${\IZ}_n$ charge conditions for integral $p \ge N$. 
This conjectural decomposition generalizes 
the decomposition formula \eqref{eq:decomp_ch} 
of the generating functions of $N$-tuples of $n$-coloured Young diagrams 
for $p \to \infty$ (or for a generic central charge), and gives 
the $\mathcal{W}^{\, para}_{N, n}$ $(p,p+n)$-minimal model characters 
(branching functions of the coset factor in $\mathcal{A}(N,n;p)$). 
When $p=N$, the central charge of 
the $\mathcal{W}^{\, para}_{N, n}$ $(N,N+n)$-minimal model is 
vanished, and in Remark \ref{rm:sp_reduce} 
the conjectural formula is indeed shown to yield 
the formula \eqref{wzw_ch_comb_c} which gives 
the $\widehat{\mathfrak{sl}}(n)_N$ WZW characters. 

In \cite{Foda:2019msm} we also introduced the $SU(N)$ 
\textit{Burge-reduced instanton partition functions} on ${\IC}^2/{\IZ}_n$ 
with $2N$ (anti-)fundamental hypermultiplets, where the Burge conditions and 
the ${\IZ}_n$ charge conditions for $p = N$ were imposed. 
We then conjectured that they give the specific integrable $\widehat{\mathfrak{sl}}(n)_N$ WZW 4-point conformal blocks 
in \cite{Knizhnik:1984nr}.  
Similarly to the conjectural decomposition \eqref{eq:conj_minimal} 
of the Burge-reduced generating functions, 
the Burge-reduced instanton partition functions for integral $p \ge N$ 
are also expected to be decomposed into $\mathcal{W}^{\, para}_{N, n}$ $(p,p+n)$-minimal model conformal blocks and $\widehat{\mathfrak{sl}}(n)_N$ WZW conformal blocks (see \cite{Belavin:2012aa} in the case of $(N,n)=(2,2)$ 
with a generic central charge). 
It would be interesting to pursue this direction as was discussed in \cite{Belavin:2015ria} when $n=1$.


\section*{Acknowledgements}

The author would like to thank O Foda, N Macleod and T Welsh 
for the stimulating collaboration on \cite{Foda:2019msm} and useful discussions, 
and the Australian Research Council for support of this work.


\appendix


\renewcommand{\dimaff}{{M}}    
\renewcommand{\levelaff}{{m}}    
\section{Some string functions}\label{app:string_fn}

\textit{\noindent
In this appendix we summarize some normalized $\slchap{\dimaff}$ 
string functions of level-$\levelaff$.} 

The normalized string function $\hat{c}^{\,\Lambda}_{\,\gamma(\bel)}(\fq)$, 
for a dominant highest-weight 
$\Lambda=[d_0,d_1,\ldots,d_{\dimaff-1}] \in P^{+}_{\dimaff, \levelaff}$ and 
a maximal-weight $\gamma(\bel)=[\gamma_0,\gamma_1,\ldots,\gamma_{\dimaff-1}] \in P_{\dimaff, \levelaff}$, is obtained from 
the $\widehat{\mathfrak{sl}}(\dimaff)_{\levelaff}$ WZW character 
in \eqref{wzw_ch} as 
(see eqns.~\eqref{wzw_string} and \eqref{gen_string_fn_Nn} with the normalization by the central charge),
\begin{align}
\chi_{\Lambda}^{\widehat{\mathfrak{sl}}(\dimaff)_{\levelaff}}
(\fq, \hat{\bft}) = \fq^{\, \frac{1}{2\levelaff}\, |\gamma(\bel)|^2}\, 
\sum_{\bel \in {\IZ}^{\dimaff-1}}
\hat{c}^{\, \Lambda}_{\, \gamma(\bel)}(\fq)\, 
\prod_{i=1}^{\dimaff-1} \hat{\ft}_{i}^{\, \gamma_i(\bel)},
\label{wzw_string_app}
\end{align}
where $\gamma_i=\gamma_i(\bel)=d_i+\ell_{i-1}-2\ell_i+\ell_{i+1}$ with 
$\ell_{\dimaff}=\ell_0=0$, $\ell_{-1}=\ell_{\dimaff-1}$. 
The WZW characters can be computed by the Weyl-Kac character 
formula \cite{kac.book.1990} 
(see also \cite[Appendix B.2]{Foda:2015bsa} and \cite[Appendix A.8]{Foda:2019msm}),
\begin{align}
\chi_{\Lambda}^{\widehat{\mathfrak{sl}}(\dimaff)_{\levelaff}}(\fq, \hat{\bft}) =
\frac{\mathcal{N}_{\Lambda}\left( \fq, \hat{\bft} \right)\, \fq^{\, h_{\Lambda}}}
{\left( \fq;\fq \right)_{\infty}^{\dimaff-1}\, 
\prod_{1 \le i < j \le \dimaff}
\left(\hat{\ft}_{i-1}\hat{\ft}_{j}/\hat{\ft}_{i}\hat{\ft}_{j-1};\fq \right)_{\infty}
\left( \fq \, \hat{\ft}_{i}\hat{\ft}_{j-1}/\hat{\ft}_{i-1}\hat{\ft}_{j};\fq \right)_{\infty}}\,
\prod_{i=1}^{\dimaff-1} \hat{\ft}_{i}^{\, d_i},
\label{wzw_ch_formula}
\end{align}
with
\begin{align}
\mathcal{N}_{\Lambda}(\fq, \hat{\bft}) =
\mathop{\sum_{(k_1,\ldots,k_{\dimaff}) \in {\IZ}^{\dimaff}}}\limits_{k_1 + \cdots + k_{\dimaff}=0}
\det_{1 \le i, j \le \dimaff}
\ll \ll \hat{\ft}_i/\hat{\ft}_{i-1} \rr^{\ll \dimaff+\levelaff \rr k_i-\lambda_i + i + \lambda_j - j}\, 
\fq^{\, \frac12\ll \dimaff+\levelaff\rr k_i^2+
\ll \lambda_j-j\rr k_i}\rr,
\end{align}
where $\left(\fq;\fq \right)_{\infty}=\prod_{n=1}^{\infty}(1-\fq^n)$, 
$\hat{\ft}_0=\hat{\ft}_{\dimaff}=1$ and 
$(\lambda_1,\lambda_2,\ldots)=\partit(\Lambda)$ in \eqref{partition_L}. 
Note that the string functions are invariant 
under the outer automorphisms of $\widehat{\mathfrak{sl}}(\dimaff)$ as 
\begin{align}
\hat{c}^{\,\Lambda}_{\,\gamma(\bel)}(\fq)
= \hat{c}^{\,\Lambda^{(\theta)}}_{\,\gamma(\bel)^{(\theta)}}(\fq),
\quad
\theta \in {\IZ}_{\dimaff},
\end{align}
where $\Lambda^{(\theta)}=[d_0^{(\theta)},d_1^{(\theta)},\ldots,d_{\dimaff-1}^{(\theta)}]$
with $d^{(\theta)}_i=d_{i-\theta}$, $d_{i+\dimaff}=d_i$, 
and 
$\gamma(\bel)^{(\theta)}=[\gamma_0^{(\theta)},\gamma_1^{(\theta)},\ldots,\gamma_{\dimaff-1}^{(\theta)}]$
with $\gamma^{(\theta)}_i=\gamma_{i-\theta}$, $\gamma_{i+\dimaff}=\gamma_i$.

\subsection{$\slchap{2}$}

When $\dimaff=2$, the $\slchap{2}$ string functions 
$\hat{c}^{\, [\levelaff-d, d]}_{\,[\levelaff-\gamma,\gamma]}(\fq)$ of level-$\levelaff$, 
with $d - \gamma \in 2{\IZ}$, for $[\levelaff-d, d] \in P^{+}_{2,\levelaff}$ 
and $[\levelaff-\gamma, \gamma] \in P_{2,\levelaff}$ are given by 
\cite{Distler:1989xv},
\begin{align}
\begin{split}
\hat{c}^{\, [\levelaff-d, d]}_{\,[\levelaff-\gamma,\gamma]}(\fq)&=
\frac{\fq^{\, \frac{d(d+2)}{4(m+2)}-\frac{\gamma^2}{4m} }}
{\left(\fq;\fq \right)_{\infty}^3}\,
\sum_{k_1,k_2=0}^{\infty} (-1)^{k_1 + k_2}\,
\fq^{\frac12k_1(k_1+1)+\frac12k_2(k_2+1)+(\levelaff+1)k_1k_2}
\\
&\qquad\qquad\qquad \times
\left(\fq^{\frac12(d-\gamma)k_1 + \frac12(d+\gamma)k_2}
- \fq^{\levelaff + 1 - d + \frac12(2\levelaff+2-d+\gamma)k_1 + \frac12(2\levelaff+2-d-\gamma)k_2}\right),
\label{string_fn_N2}
\end{split}
\end{align}
and satisfy 
$\hat{c}^{\, [\levelaff-d, d]}_{\,[\levelaff-\gamma,\gamma]}(\fq)=
\hat{c}^{\, [d, \levelaff-d]}_{\,[\gamma,\levelaff-\gamma]}(\fq)$.

\subsection{$\slchap{3}$}\label{app:sl3_string}

Here we summarize the $\slchap{3}$ string functions $\hat{c}^{\, \Lambda}_{\,\gamma}(\fq)$ of level-2 and 3 given in \cite{Kac:1984mq}.

The $\slchap{3}$ string functions of level-2 are
\begin{align}
\begin{split}
&
\hat{c}^{\,[2,0,0]}_{\,[2,0,0]}(\fq) - \hat{c}^{\,[2,0,0]}_{\,[0,1,1]}(\fq)=
\frac{\left( \fq^{\frac12};\fq^{\frac12} \right)_{\infty}
\left( \fq, \fq^{\frac32}, \fq^{\frac52};\fq^{\frac52} \right)_{\infty}}
{\left( \fq;\fq \right)_{\infty}^4},
\\
&
\hat{c}^{\,[2,0,0]}_{\,[0,1,1]}(\fq)= \fq^{\frac12} \, 
\frac{\left( \fq^{2};\fq^{2} \right)_{\infty}
\left( \fq^2, \fq^{8}, \fq^{10};\fq^{10} \right)_{\infty}}
{\left( \fq;\fq \right)_{\infty}^4},
\\
&
\hat{c}^{\,[0,1,1]}_{\,[0,1,1]}(\fq)= \fq^{\frac{1}{10}} \,
\frac{\left( \fq^{2};\fq^{2} \right)_{\infty}
\left( \fq^4, \fq^{6}, \fq^{10};\fq^{10} \right)_{\infty}}
{\left( \fq;\fq \right)_{\infty}^4},
\\
&
\hat{c}^{\,[0,1,1]}_{\,[0,1,1]}(\fq) - 
\hat{c}^{\,[0,1,1]}_{\,[2,0,0]}(\fq)= \fq^{\frac{1}{10}} \, 
\frac{\left( \fq^{\frac12};\fq^{\frac12} \right)_{\infty}
\left( \fq^{\frac12}, \fq^{2}, \fq^{\frac52};\fq^{\frac52} \right)_{\infty}}
{\left( \fq;\fq \right)_{\infty}^4},
\label{string_fn_N3_l2}
\end{split}
\end{align}
where 
$\left(a_1, a_2, \ldots, a_k ; \fq \right)_{\infty}=
\prod_{i=1}^{k}\left(a_i ; \fq \right)_{\infty}=
\prod_{i=1}^{k}\prod_{n=1}^{\infty}(1-a_i\fq^{n-1})$.

The $\slchap{3}$ string functions of level-3 are
\begin{align}
\begin{split}
&
\hat{c}^{\,[3,0,0]}_{\,[3,0,0]}(\fq) - \hat{c}^{\,[3,0,0]}_{\,[0,3,0]}(\fq)=
\frac{1}{\left( \fq;\fq \right)_{\infty}
\left( \fq^3;\fq^3 \right)_{\infty}},
\qquad
\hat{c}^{\,[1,1,1]}_{\,[1,1,1]}(\fq)= \fq^{\frac16}\,
\frac{\left( \fq^2;\fq^2 \right)_{\infty}^3
\left( \fq^3;\fq^3 \right)_{\infty}^2}
{\left( \fq;\fq \right)_{\infty}^6
\left( \fq^6;\fq^6 \right)_{\infty}},
\\
&
\hat{c}^{\,[2,1,0]}_{\,[2,1,0]}(\fq)+\hat{c}^{\,[2,1,0]}_{\,[0,2,1]}(\fq)+
\hat{c}^{\,[2,1,0]}_{\,[1,0,2]}(\fq)=
\frac{\fq^{\frac19}}{\left( \fq;\fq \right)_{\infty}
\left( \fq^{\frac13};\fq^{\frac13} \right)_{\infty}},
\\
&
\hat{c}^{\,[3,0,0]}_{\,[3,0,0]}(\fq)-3\, \hat{c}^{\,[3,0,0]}_{\,[1,1,1]}(\fq)
+2\, \hat{c}^{\,[3,0,0]}_{\,[0,3,0]}(\fq)+\hat{c}^{\,[1,1,1]}_{\,[1,1,1]}(\fq)
-\hat{c}^{\,[1,1,1]}_{\,[3,0,0]}(\fq)=
\frac{\left( \fq^{\frac12};\fq^{\frac12} \right)_{\infty}^3
\left( \fq^{\frac13};\fq^{\frac13} \right)_{\infty}^2}
{\left( \fq;\fq \right)_{\infty}^6
\left( \fq^{\frac16};\fq^{\frac16} \right)_{\infty}},
\label{string_fn_N3_l3}
\end{split}
\end{align}
where $\hat{c}^{\,[2,1,0]}_{\,[2,1,0]}(\fq) \in \fq^{\frac19}\,{\IZ}[[\fq]]$, 
$\hat{c}^{\,[2,1,0]}_{\,[0,2,1]}(\fq) \in \fq^{\frac49}\,{\IZ}[[\fq]]$, 
$\hat{c}^{\,[2,1,0]}_{\,[1,0,2]}(\fq) \in \fq^{\frac79}\,{\IZ}[[\fq]]$, 
$\hat{c}^{\,[3,0,0]}_{\,[1,1,1]}(\fq)\in \fq^{\frac23}\,{\IZ}[[\fq]]$ and 
$\hat{c}^{\,[1,1,1]}_{\,[3,0,0]}(\fq)\in \fq^{\frac12}\,{\IZ}[[\fq]]$.

\section{Examples of dual dominant integral weights}\label{app:hw_ex}

\textit{\noindent 
Here we provide some examples of 
the dominant integral weights $\bN_{\bel}^{(f)}
=\left[N_0,N_1,\ldots,N_{n-1}\right]\in P^{+}_{n,N}$ 
of $\widehat{\mathfrak{sl}}(n)_N$ in \eqref{dw_dual}, 
which are labelled by a non-negative integer $f< \mathrm{max}\{N, n\}$ 
and dominant integral weights $\bel=\left[\ell_0,\ell_1,\ldots,\ell_{N-1}\right]
\in P^{+}_{N,n}$ of $\widehat{\mathfrak{sl}}(N)_n$.}

For $(N,n)=(2,2)$,
\begin{align}
\bN_{[2,0]}^{(0)}=[2,0],\quad
\bN_{[0,2]}^{(0)}=[0,2],\quad
\bN_{[1,1]}^{(1)}=[1,1].
\end{align}
For $(N,n)=(2,3)$,
\begin{align}
\begin{split}
\bN_{[3,0]}^{(0)}&=[2,0,0],\ \bN_{[1,2]}^{(0)}=[0,1,1],\ 
\bN_{[3,0]}^{(2)}=[0,2,0],\ \bN_{[1,2]}^{(2)}=[1,0,1],
\\ 
\bN_{[2,1]}^{(1)}&=[1,1,0],\ \bN_{[0,3]}^{(1)}=[0,0,2].
\end{split}
\end{align}
For $(N,n)=(2,4)$,
\begin{align}
\begin{split}
&
\bN_{[4,0]}^{(0)}=[2,0,0,0],\ \bN_{[2,2]}^{(0)}=[0,1,0,1],\ \bN_{[0,4]}^{(0)}=[0,0,2,0],
\\ 
&
\bN_{[4,0]}^{(2)}=[0,2,0,0],\ \bN_{[2,2]}^{(2)}=[1,0,1,0],\ \bN_{[0,4]}^{(2)}=[0,0,0,2],
\\
&
\bN_{[3,1]}^{(1)}=[1,1,0,0],\ \bN_{[1,3]}^{(1)}=[0,0,1,1],\ 
\bN_{[3,1]}^{(3)}=[0,1,1,0],\ \bN_{[1,3]}^{(3)}=[1,0,0,1].
\end{split}
\end{align}
For $(N,n)=(3,2)$,
\begin{align}
\bN_{[2,0,0]}^{(0)}=\bN_{[0,2,0]}^{(2)}=[3,0],\quad
\bN_{[0,1,1]}^{(0)}=\bN_{[1,0,1]}^{(2)}=[1,2],\quad
\bN_{[1,1,0]}^{(1)}=[2,1],\quad
\bN_{[0,0,2]}^{(1)}=[0,3].
\end{align} 
For $(N,n)=(3,3)$,
\begin{align}
\begin{split}
&
\bN_{[3,0,0]}^{(0)}=[3,0,0],\quad
\bN_{[1,1,1]}^{(0)}=[1,1,1],\quad
\bN_{[0,0,3]}^{(0)}=[0,3,0],\quad
\bN_{[0,3,0]}^{(0)}=[0,0,3],
\\
&
\bN_{[2,1,0]}^{(1)}=[2,1,0],\quad
\bN_{[1,0,2]}^{(1)}=[0,2,1],\quad
\bN_{[0,2,1]}^{(1)}=[1,0,2],
\\
&
\bN_{[1,2,0]}^{(2)}=[2,0,1],\quad
\bN_{[2,0,1]}^{(2)}=[1,2,0],\quad
\bN_{[0,1,2]}^{(2)}=[0,1,2].
\end{split}
\end{align}
For $(N,n)=(4,2)$,
\begin{align}
\begin{split}
&
\bN_{[2,0,0,0]}^{(0)}=\bN_{[0,2,0,0]}^{(2)}=[4,0],\quad
\bN_{[0,1,0,1]}^{(0)}=\bN_{[1,0,1,0]}^{(2)}=[2,2],\quad
\bN_{[0,0,2,0]}^{(0)}=\bN_{[0,0,0,2]}^{(2)}=[0,4],
\\
&
\bN_{[1,1,0,0]}^{(1)}=\bN_{[0,1,1,0]}^{(3)}=[3,1],\quad
\bN_{[0,0,1,1]}^{(1)}=\bN_{[1,0,0,1]}^{(3)}=[1,3].
\end{split}
\end{align}

\section{More examples of Burge-reduced generating functions}\label{app:add_examples}

\textit{\noindent
In this appendix, in addition to the examples in Section \ref{sec:examples}, 
we give some more examples of 
the $SU(N)$ Burge-reduced generating functions of $n$-coloured Young diagrams 
in the cases of $(N,n,p)=(2,3,3), (2,4,4), (3,2,4)$ and $(4,2,5)$ and 
check Conjecture \ref{conj:suN_red_ch}.}

\subsection{$(N,n,p)=(2,3,3)$}

Consider the case of $(N,n)=(2,3)$ and $p=3$. 
The $\mathcal{W}^{\, para}_{2, 3}$ $(3,6)$-minimal model has 
central charge $c\left(\mathcal{W}^{\, para}_{2, 3}\right)=4/5$. 
The Burge-reduced generating functions for 
$\bN=[2,0,0]$ and $\bfc=[2,0,0]$, $[0,1,1]$ are obtained as
\begin{align}
\begin{split}
\widehat{X}_{(0,0);(0,0)}^{\,[2,1],[5,1]}(\fq)&=
1+2 \fq+11 \fq^2+32 \fq^3+97 \fq^{4}+246 \fq^{5}+610 \fq^{6}+1388 \fq^{7}+3067 \fq^{8}+\cdots \, ,
\\
\widehat{X}_{(0,0);(-1,-1)}^{\,[2,1],[5,1]}(\fq)&=
\fq^{\frac13}+5 \fq^{\frac43}+18 \fq^{\frac73}+56 \fq^{\frac{10}{3}}+154 \fq^{\frac{13}{3}}+389 \fq^{\frac{16}{3}}+922 \fq^{\frac{19}{3}}+2072 \fq^{\frac{22}{3}}+\cdots \, ,
\end{split}
\end{align}
and using the $\widehat{\mathfrak{sl}}(3)$ string functions \eqref{string_fn_N3_l2} of level-2 
with $\hat{a}^{\, \bN}_{\,\bfc}(\fq)=
\fq^{\,\frac{1}{6}(\fc_1^2+\fc_2^2+\fc_1\fc_2)-\frac13(\fc_1+\fc_2)}\,\hat{c}^{\, \bN}_{\,\bfc}(\fq)$, 
from the formula \eqref{eq:conj_minimal_c} 
we obtain the $\mathcal{W}^{\, para}_{2, 3}$ $(3,6)$-minimal model characters
\begin{align}
\begin{split}
C_{\,[3,0]}^{\, [2,1],[5,1]}(\fq)&=
1+\fq^2+\fq^3+2 \fq^{4}+2 \fq^{5}+4 \fq^{6}+4 \fq^{7}+7 \fq^{8}+\cdots\,,
\\
C_{\,[1,2]}^{\, [2,1],[5,1]}(\fq)&=
\fq^{\frac75}+\fq^{\frac{12}{5}}+2 \fq^{\frac{17}{5}}+2 \fq^{\frac{22}{5}}+4 \fq^{\frac{27}{5}}+5 \fq^{\frac{32}{5}}+8 \fq^{\frac{37}{5}}+\cdots\,.
\end{split}
\end{align}
Similarly, the Burge-reduced generating functions for 
$\bN=[0,1,1]$ and $\bfc=[2,0,0]$, $[0,1,1]$,
\begin{align}
\begin{split}
\widehat{X}_{(2,1);(1,1)}^{\,[2,1],[4,2]}(\fq)&=
2 \fq^{\frac23}+10 \fq^{\frac53}+36 \fq^{\frac83}+110 \fq^{\frac{11}{3}}+300 \fq^{\frac{14}{3}}+752 \fq^{\frac{17}{3}}+1770 \fq^{\frac{20}{3}}+3956 \fq^{\frac{23}{3}}+\cdots \, ,
\\
\widehat{X}_{(2,1);(0,0)}^{\,[2,1],[4,2]}(\fq)&=
1+5 \fq+20 \fq^2+65 \fq^3+185 \fq^{4}+481 \fq^{5}+1165 \fq^{6}+2665 \fq^{7}+5822 \fq^{8}+\cdots \, ,
\end{split}
\end{align}
give
\begin{align}
\begin{split}
C_{\,[2,1]}^{\, [2,1],[4,2]}(\fq)&=
\fq^{\frac{1}{15}}\,
(1+\fq+2 \fq^{2}+3 \fq^{3}+4 \fq^{4}+6 \fq^{5}+9 \fq^{6}+12 \fq^{7}+17 \fq^{8}+\cdots)\,,
\\
C_{\,[0,3]}^{\, [2,1],[4,2]}(\fq)&=
\fq^{\frac53}+\fq^{\frac83}+2 \fq^{\frac{11}{3}}+3 \fq^{\frac{14}{3}}+4 \fq^{\frac{17}{3}}+6 \fq^{\frac{20}{3}}+9 \fq^{\frac{23}{3}}+\cdots\,.
\end{split}
\end{align}

\subsection{$(N,n,p)=(2,4,4)$}

When $(N,n)=(2,4)$, the $\mathcal{W}^{\, para}_{2, 4}$ is known as 
the $S_3$ parafermion algebra \cite{Fateev:1985ig} and 
also discussed in the context of the AGT correspondence 
in \cite{Wyllard:2011mn, Alfimov:2011ju}. 
Here we consider the case of $p=4$, and 
the $\mathcal{W}^{\, para}_{2, 4}$ $(4,8)$-minimal model has 
central charge $c\left(\mathcal{W}^{\, para}_{2, 4}\right)=5/4$. 
The Burge-reduced generating functions for 
$\bN=[2,0,0,0]$ and $\bfc=[2,0,0,0], [0,1,0,1]$, $[0,0,2,0]$ are 
obtained as
\begin{align}
\begin{split}
\widehat{X}_{(0,0);(0,0,0)}^{\,[3,1],[7,1]}(\fq)&=
1+3 \fq+19 \fq^2+72 \fq^{3}+272 \fq^{4}+877 \fq^{5}+2680 \fq^{6}+7546 \fq^{7}+\cdots \, ,
\\
\widehat{X}_{(0,0);(-1,-1,-1)}^{\,[3,1],[7,1]}(\fq)&=
\fq^{\frac14}+7 \fq^{\frac54}+34 \fq^{\frac94}+137 \fq^{\frac{13}{4}}+481 \fq^{\frac{17}{4}}+1528 \fq^{\frac{21}{4}}+4490 \fq^{\frac{25}{4}}+\cdots \, ,
\\
\widehat{X}_{(0,0);(-1,-2,-1)}^{\,[3,1],[7,1]}(\fq)&=
2 \fq+14 \fq^2+66 \fq^{3}+252 \fq^{4}+852 \fq^{5}+2614 \fq^{6}+7460 \fq^{7}+\cdots \, ,
\end{split}
\end{align}
and by the formula \eqref{eq:conj_minimal_c} 
with $\hat{a}^{\, \bN}_{\,\bfc}(\fq)=
\fq^{\,\frac{1}{4}|\bfc|^2-w_{\bfc}}\,\hat{c}^{\, \bN}_{\,\bfc}(\fq)$ 
we obtain the $\mathcal{W}^{\, para}_{2, 4}$ $(4,8)$-minimal model characters
\begin{align}
\begin{split}
C_{\,[4,0]}^{\, [3,1],[7,1]}(\fq)&=
1+\fq^2+\fq^{3}+3 \fq^{4}+3 \fq^{5}+7 \fq^{6}+8 \fq^{7}+\cdots\,,
\\
C_{\,[2,2]}^{\, [3,1],[7,1]}(\fq)&=
\fq^{\frac43}+\fq^{\frac73}+3 \fq^{\frac{10}{3}}+4 \fq^{\frac{13}{3}}+8 \fq^{\frac{16}{3}}+11 \fq^{\frac{19}{3}}+\cdots\,,
\\
C_{\,[0,4]}^{\, [3,1],[7,1]}(\fq)&=
\fq^{3}+\fq^{4}+3 \fq^{5}+4 \fq^{6}+7 \fq^{7}+\cdots\,.
\end{split}
\end{align}
The Burge-reduced generating functions for 
$\bN=[0,1,0,1]$ and $\bfc=[2,0,0,0], [0,1,0,1]$, $[0,0,2,0]$,
\begin{align}
\begin{split}
\widehat{X}_{(3,1);(1,1,1)}^{\,[3,1],[5,3]}(\fq)&=
3 \fq^{\frac34}+21 \fq^{\frac74}+105 \fq^{\frac{11}{4}}+419 \fq^{\frac{15}{4}}+1469 \fq^{\frac{19}{4}}+4636 \fq^{\frac{23}{4}}+13544 \fq^{\frac{27}{4}}+\cdots \, ,
\\
\widehat{X}_{(3,1);(0,0,0)}^{\,[3,1],[5,3]}(\fq)&=
1+9 \fq+50 \fq^2+217 \fq^{3}+803 \fq^{4}+2651 \fq^{5}+8019 \fq^{6}+22618 \fq^{7}+\cdots \, ,
\\
\widehat{X}_{(3,1);(0,-1,0)}^{\,[3,1],[5,3]}(\fq)&=
4 \fq^{\frac34}+22 \fq^{\frac74}+110 \fq^{\frac{11}{4}}+426 \fq^{\frac{15}{4}}+1490 \fq^{\frac{19}{4}}+4666 \fq^{\frac{23}{4}}+13616 \fq^{\frac{27}{4}}+\cdots \, ,
\end{split}
\end{align}
give
\begin{align}
\begin{split}
C_{\,[4,0]}^{\, [3,1],[5,3]}(\fq)&=
\fq^{\frac34}+\fq^{\frac74}+3 \fq^{\frac{11}{4}}+4 \fq^{\frac{15}{4}}+8 \fq^{\frac{19}{4}}+11 \fq^{\frac{23}{4}}+19 \fq^{\frac{27}{4}}+\cdots\,,
\\
C_{\,[2,2]}^{\, [3,1],[5,3]}(\fq)&=
\fq^{\frac{1}{12}}\,
(1+\fq+3 \fq^{2}+5 \fq^{3}+10 \fq^{4}+15 \fq^{5}+26 \fq^{6}+\cdots)\,,
\\
C_{\,[0,4]}^{\, [3,1],[5,3]}(\fq)&=
\fq^{\frac74}+2 \fq^{\frac{11}{4}}+3 \fq^{\frac{15}{4}}+6 \fq^{\frac{19}{4}}+10 \fq^{\frac{23}{4}}+16 \fq^{\frac{27}{4}}+\cdots\,.
\end{split}
\end{align}
The Burge-reduced generating functions for 
$\bN=[1,1,0,0]$ and $\bfc=[1,1,0,0], [0,0,1,1]$,
\begin{align}
\begin{split}
\widehat{X}_{(1,0);(0,0,0)}^{\,[3,1],[6,2]}(\fq)&=
1+7 \fq+37 \fq^2+157 \fq^{3}+575 \fq^{4}+1889 \fq^{5}+5704 \fq^{6}+16081 \fq^{7}+\cdots \, ,
\\
\widehat{X}_{(1,0);(0,-1,-1)}^{\,[3,1],[6,2]}(\fq)&=
2 \fq^{\frac12}+15 \fq^{\frac32}+74 \fq^{\frac52}+297 \fq^{\frac72}+1039 \fq^{\frac92}+3284 \fq^{\frac{11}{2}}+9598 \fq^{\frac{13}{2}}+\cdots \, ,
\end{split}
\end{align}
give
\begin{align}
\begin{split}
C_{\,[3,1]}^{\, [3,1],[6,2]}(\fq)&=
\fq^{\frac{1}{16}}\,
(1+\fq+2 \fq^{2}+4 \fq^{3}+7 \fq^{4}+11 \fq^{5}+18 \fq^{6}+\cdots)\,,
\\
C_{\,[1,3]}^{\, [3,1],[6,2]}(\fq)&=
\fq^{\frac{25}{16}}+2 \fq^{\frac{41}{16}}+4 \fq^{\frac{57}{16}}+7 \fq^{\frac{73}{16}}+12 \fq^{\frac{89}{16}}+19 \fq^{\frac{105}{16}}+\cdots\,.
\end{split}
\end{align}

\subsection{$(N,n,p)=(3,2,4)$}

Consider the case of $(N,n)=(3,2)$ and $p=4$. 
The $\mathcal{W}^{\, para}_{3, 2}$ $(4,6)$-minimal model for $p=4$ has 
central charge $c\left(\mathcal{W}^{\, para}_{3, 2}\right)=6/5$. 
The Burge-reduced generating functions for 
$\bN=[3,0]$ and $\bfc=[3,0], [1,2]$ are obtained as
\begin{align}
\begin{split}
\widehat{X}_{(0,0,0);(0)}^{\,[2,1,1],[2,3,1]}(\fq)&=
1+3 \fq+11 \fq^2+30 \fq^3+77 \fq^4+176 \fq^{5}+385 \fq^{6}+792 \fq^{7}+1575 \fq^{8}+\cdots \, ,
\\
\widehat{X}_{(0,0,0);(-1)}^{\,[2,1,1],[2,3,1]}(\fq)&=
2 \fq^{\frac12}+7 \fq^{\frac32}+22 \fq^{\frac52}+56 \fq^{\frac72}+135 \fq^{\frac92}+297 \fq^{\frac{11}{2}}+627 \fq^{\frac{13}{2}}+1255 \fq^{\frac{15}{2}}+\cdots \, ,
\end{split}
\end{align}
and using the $\widehat{\mathfrak{sl}}(2)$ string functions \eqref{string_fn_N2} 
of level-$3$ with $\hat{a}^{\,[N_0,N_1]}_{\,[\fc_0,\fc_1]}(\fq)=
\fq^{\,\frac{1}{12}\fc_1(\fc_1-3)}\,\hat{c}^{\, [N_0,N_1]}_{\,[\fc_0,\fc_1]}(\fq)$, 
from the formula \eqref{eq:conj_minimal_c} we obtain 
the $\mathcal{W}^{\, para}_{3, 2}$ $(4,6)$-minimal model characters
\begin{align}
\begin{split}
C_{\,[0,2,0]}^{\, [2,1,1],[2,3,1]}(\fq)&=
1+\fq+2 \fq^2+3 \fq^3+6 \fq^4+9 \fq^{5}+15 \fq^{6}+22 \fq^{7}+35 \fq^{8}+\cdots\,,
\\
C_{\,[1,0,1]}^{\, [2,1,1],[2,3,1]}(\fq)&=
\fq^{\frac35}+2 \fq^{\frac85}+4 \fq^{\frac{13}{5}}+7 \fq^{\frac{18}{5}}+12 \fq^{\frac{23}{5}}+19 \fq^{\frac{28}{5}}+31 \fq^{\frac{33}{5}}+46 \fq^{\frac{38}{5}}+\cdots\,.
\end{split}
\end{align}
The Burge-reduced generating functions for 
$\bN=[1,2]$ and $\bfc=[3,0], [1,2]$,
\begin{align}
\begin{split}
\widehat{X}_{(1,1,0);(1)}^{\,[2,1,1],[3,1,2]}(\fq)&=
\fq^{\frac12}+5 \fq^{\frac32}+15 \fq^{\frac52}+42 \fq^{\frac72}+101 \fq^{\frac92}+231 \fq^{\frac{11}{2}}+490 \fq^{\frac{13}{2}}+1002 \fq^{\frac{15}{2}}+\cdots \, ,
\\
\widehat{X}_{(1,1,0);(0)}^{\,[2,1,1],[3,1,2]}(\fq)&=
1+3 \fq+11 \fq^2+30 \fq^3+77 \fq^4+176 \fq^{5}+385 \fq^{6}+792 \fq^{7}+1575 \fq^{8}+\cdots \, ,
\end{split}
\end{align}
give
\begin{align}
\begin{split}
C_{\,[0,2,0]}^{\, [2,1,1],[3,1,2]}(\fq)&=
\fq^{\frac32}+2 \fq^{\frac52}+4 \fq^{\frac72}+6 \fq^{\frac92}+11 \fq^{\frac{11}{2}}+16 \fq^{\frac{13}{2}}+26 \fq^{\frac{15}{2}}+\cdots\,,
\\
C_{\,[1,0,1]}^{\, [2,1,1],[3,1,2]}(\fq)&=
\fq^{\frac{1}{10}}\,
(1+\fq+3 \fq^{2}+5 \fq^{3}+9 \fq^{4}+14 \fq^{5}+23 \fq^{6}+35 \fq^{7}+\cdots)\,.
\end{split}
\end{align}
The Burge-reduced generating functions for 
$\bN=[2,1]$ and $\bfc=[2,1], [0,3]$,
\begin{align}
\begin{split}
\widehat{X}_{(1,0,0);(0)}^{\,[1,1,2],[2,2,2]}(\fq)&=
1+5 \fq+17 \fq^2+48 \fq^3+120 \fq^4+277 \fq^{5}+600 \fq^{6}+1237 \fq^{7}+2448 \fq^{8}+\cdots \, ,
\\
\widehat{X}_{(1,0,0);(-1)}^{\,[1,1,2],[2,2,2]}(\fq)&=
2 \fq^{\frac12}+8 \fq^{\frac32}+24 \fq^{\frac52}+66 \fq^{\frac72}+160 \fq^{\frac92}+360 \fq^{\frac{11}{2}}+768 \fq^{\frac{13}{2}}+1560 \fq^{\frac{15}{2}}+\cdots \, ,
\end{split}
\end{align}
give
\begin{align}
\begin{split}
C_{\,[1,1,0]}^{\, [1,1,2],[2,2,2]}(\fq)&=
\fq^{\frac{1}{10}}\,
(1+2 \fq+4 \fq^{2}+8 \fq^{3}+13 \fq^{4}+22 \fq^{5}+35 \fq^{6}+54 \fq^{7}+\cdots)\,,
\\
C_{\,[0,0,2]}^{\, [1,1,2],[2,2,2]}(\fq)&=
\fq^{\frac12}+2 \fq^{\frac32}+3 \fq^{\frac52}+6 \fq^{\frac72}+10 \fq^{\frac92}+16 \fq^{\frac{11}{2}}+26 \fq^{\frac{13}{2}}+40 \fq^{\frac{15}{2}}+\cdots\,.
\end{split}
\end{align}

\subsection{$(N,n)=(4,2,5)$}

Consider the case of $(N,n)=(4,2)$ for $p=5$. 
The $\mathcal{W}^{\, para}_{4, 2}$ $(5,7)$-minimal model for $p=5$ has 
central charge $c\left(\mathcal{W}^{\, para}_{4, 2}\right)=11/7$. 
The Burge-reduced generating functions for 
$\bN=[4,0]$ and $\bfc=[4,0], [2,2], [0,4]$ are obtained as
\begin{align}
\begin{split}
\widehat{X}_{(0,0,0,0);(0)}^{\,[2,1,1,1],[2,3,1,1]}(\fq)&=
1+3 \fq+11 \fq^2+34 \fq^3+93 \fq^4+234 \fq^{5}+552 \fq^{6}+\cdots\, ,
\\
\widehat{X}_{(0,0,0,0);(-1)}^{\,[2,1,1,1],[2,3,1,1]}(\fq)&=
2 \fq^{\frac12}+7 \fq^{\frac32}+25 \fq^{\frac52}+70 \fq^{\frac72}+185 \fq^{\frac92}+441 \fq^{\frac{11}{2}}+\cdots\, ,
\\
\widehat{X}_{(0,0,0,0);(-2)}^{\,[2,1,1,1],[2,3,1,1]}(\fq)&=
2 \fq+9 \fq^2+31 \fq^3+88 \fq^4+227 \fq^{5}+541 \fq^{6}+\cdots\, ,
\end{split}
\end{align}
and using the $\widehat{\mathfrak{sl}}(2)$ string functions \eqref{string_fn_N2} 
of level-$4$ with $\hat{a}^{\,[N_0,N_1]}_{\,[\fc_0,\fc_1]}(\fq)=
\fq^{\,\frac{1}{16}\fc_1(\fc_1-4)}\,\hat{c}^{\, [N_0,N_1]}_{\,[\fc_0,\fc_1]}(\fq)$, 
from the formula \eqref{eq:conj_minimal_c} we obtain 
the $\mathcal{W}^{\, para}_{4, 2}$ $(5,7)$-minimal model characters
\begin{align}
\begin{split}
C_{\,[0,2,0,0]}^{\, [2,1,1,1],[2,3,1,1]}(\fq)&=
1+\fq+2 \fq^2+4 \fq^3+7 \fq^4+12 \fq^{5}+21 \fq^{6}+\cdots\,,
\\
C_{\,[1,0,1,0]}^{\, [2,1,1,1],[2,3,1,1]}(\fq)&=
\fq^{\frac23}+2 \fq^{\frac53}+5 \fq^{\frac83}+9 \fq^{\frac{11}{3}}+18 \fq^{\frac{14}{3}}+30 \fq^{\frac{17}{3}}+\cdots\,,
\\
C_{\,[0,0,0,2]}^{\, [2,1,1,1],[2,3,1,1]}(\fq)&=
\fq^2+2 \fq^3+5 \fq^4+9 \fq^{5}+17 \fq^{6}+\cdots\,.
\end{split}
\end{align}
The Burge-reduced generating functions for 
$\bN=[2,2]$ and $\bfc=[4,0], [2,2], [0,4]$,
\begin{align}
\begin{split}
\widehat{X}_{(1,1,0,0);(1)}^{\,[2,1,1,1],[3,1,2,1]}(\fq)&=
\fq^{\frac12}+5 \fq^{\frac32}+18 \fq^{\frac52}+55 \fq^{\frac72}+149 \fq^{\frac92}+371 \fq^{\frac{11}{2}}+\cdots\, ,
\\
\widehat{X}_{(1,1,0,0);(0)}^{\,[2,1,1,1],[3,1,2,1]}(\fq)&=
1+3 \fq+14 \fq^2+41 \fq^3+119 \fq^4+295 \fq^{5}+706 \fq^{6}+\cdots\, ,
\\
\widehat{X}_{(1,1,0,0);(-1)}^{\,[2,1,1,1],[3,1,2,1]}(\fq)&=
\fq^{\frac12}+5 \fq^{\frac32}+18 \fq^{\frac52}+55 \fq^{\frac72}+149 \fq^{\frac92}+371 \fq^{\frac{11}{2}}+\cdots\, ,
\end{split}
\end{align}
give
\begin{align}
\begin{split}
C_{\,[0,2,0,0]}^{\, [2,1,1,1],[3,1,2,1]}(\fq)&=
\fq^{\frac32}+2 \fq^{\frac52}+5 \fq^{\frac72}+8 \fq^{\frac92}+16 \fq^{\frac{11}{2}}+\cdots\,,
\\
C_{\,[1,0,1,0]}^{\, [2,1,1,1],[3,1,2,1]}(\fq)&=
\fq^{\frac16}\,
(1+\fq+4 \fq^{2}+7 \fq^{3}+15 \fq^{4}+25 \fq^{5}+\cdots)\,,
\\
C_{\,[0,0,0,2]}^{\, [2,1,1,1],[3,1,2,1]}(\fq)&=
\fq^{\frac32}+2 \fq^{\frac52}+5 \fq^{\frac72}+8 \fq^{\frac92}+16 \fq^{\frac{11}{2}}+\cdots\,.
\end{split}
\end{align}


\end{document}